\documentclass{aa}

\usepackage{amssymb,amsmath}
\usepackage{subfig}
\usepackage{graphicx}
\usepackage{xcolor}
\usepackage{hyperref}
\usepackage{threeparttable}
\usepackage{multirow}
\usepackage{arydshln}
\definecolor{yaleblue}{rgb}{0.1,0.3,0.9}
\definecolor{lava}{rgb}{0.81, 0.06, 0.13}
\definecolor{forestgreen}{rgb}{0.0, 0.27, 0.13}
\hypersetup{colorlinks=true, linkcolor=lava, urlcolor=forestgreen, citecolor=yaleblue}
\usepackage{listings}
\usepackage{color}
\usepackage{soul}
\setstcolor{orange}

\graphicspath{{figures/}}

\newcommand{\orcid}[1]{\href{https://orcid.org/#1}{\includegraphics[width=10pt]{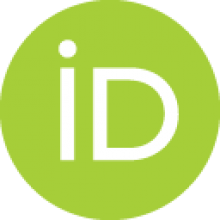}}}

\begin{document}

\definecolor{dkgreen}{rgb}{0,0.6,0}
\definecolor{gray}{rgb}{0.5,0.5,0.5}
\definecolor{mauve}{rgb}{0.58,0,0.82}

\lstset{frame=tb,
  	language=Matlab,
  	aboveskip=3mm,
  	belowskip=3mm,
  	showstringspaces=false,
  	columns=flexible,
  	basicstyle={\small\ttfamily},
  	numbers=none,
  	numberstyle=\tiny\color{gray},
 	keywordstyle=\color{blue},
	commentstyle=\color{dkgreen},
  	stringstyle=\color{mauve},
  	breaklines=true,
  	breakatwhitespace=true
  	tabsize=3
}

\title{Moonlit sky polarization patterns from Cerro Paranal}
\author{B. Pereira \inst{1}\orcid{0009-0005-6455-1228}
\and S.  González-Gaitán \inst{2,3}\orcid{0000-0001-9541-0317}
\and A. M. Mourão \inst{1}
\and J. Rino-Silvestre \inst{1}
\and A. Paulino-Afonso \inst{4} 
\and J.~P. Anderson \inst{3,5}\orcid{0000-0003-0227-3451}
\and A. Cikota \inst{6}
\and A. Morales-Garoffolo \inst{7}\orcid{0000-0001-8830-7063}}
\institute{
Department of Physics and CENTRA- Center for Astrophysics and Gravitation,  Instituto Superior Técnico, University of Lisbon, Lisbon, Portugal
\and
Instituto de Astrof\'isica e Ci\^encias do Espaço, Faculdade de Ci\^encias, Universidade de Lisboa, Ed. C8, Campo Grande, 1749-016 Lisbon, Portugal\\
    \email{gongsale@gmail.com}    
    \and
        European Southern Observatory, Alonso de C\'ordova 3107, Casilla 19, Santiago, Chile
\and Instituto de Astrof\'isica e Ci\^encias do Espa\c co, Universidade do Porto, CAUP, Rua das Estrelas, Porto, 4150-762, Portugal
\and Millennium Institute of Astrophysics MAS, Nuncio Monsenor Sotero Sanz 100, Off. 104, Providencia, Santiago, Chile
 \and
    Gemini Observatory/NSF’s NOIRLab, Casilla 603, La Serena, Chile
\and Department of Applied Physics, School of Engineering, University of C\'adiz, Campus of Puerto Real, E-11519 C\'adiz, Spain
}

\date{Received XX, 2025; accepted XX, 2025}

\abstract
{
We investigate the polarization patterns from the moonlit sky as observed from the European Southern Observatory at Cerro Paranal. The moonlit sky background can be significant in astronomical observations and thus be a source of contamination in polarimetric studies. Based on sky observations during full Moon with FORS2 in imaging polarimetric mode, we measure the polarization degree and intensity at different wavelengths and scattering angles from the Moon, and compare them to theoretical and phenomenological single and multiple scattering models. 
Single scattering Rayleigh models are able to reproduce the wavelength dependence of the polarization as long as strong depolarization factors that increase with wavelength are introduced. Intensity data, however, require the inclusion of single Mie scattering from larger aerosol particles. The best models that simultaneously fit polarization and intensity data, are a combination of both single scattering processes, Rayleigh and Mie, plus an unpolarized multiple scattering component. Both Mie and multiple scattering become more dominant at longer wavelengths.  
Other factors like cloud depolarization and the sunlight contribution during the twilight were also investigated. The present study underscores the importance of accounting for moonlight scattering to enhance the accuracy of polarimetric observations of astronomical targets.
}
 
\authorrunning{Pereira, Gonz\'alez-Gait\'an et al.}
\maketitle

\section{\label{sec:intro} Introduction}

The study of polarized light has been crucial in the progress of astronomy. Unique physical properties of astronomical objects can be deduced by analyzing polarized light that reaches the observer. It complements intensity-based methods like photometry and spectroscopy. For example, optical linear polarimetry has helped consolidate the unification model of active galactic nuclei \citep{Antonucci85}, it reveals asymmetries of the ejecta in supernova explosions \citep[e.g.][]{Leonard01}, magnetic field structure in galaxies \citep[e.g.][]{Scarrott91}, dust characteristics in galaxies \citep{Rino-Silvestre25} and the composition and geometry of kilonovae \citep[e.g.][]{Bulla19}. 

Polarimetric observations from ground-based telescopes present various challenges. In particular, the scattering of moonlight in Earth's atmosphere generates polarization patterns that may contaminate polarimetric measurements of astronomical targets. Light polarization by atmospheric scattering has been studied extensively throughout the years. It depends on various factors such as the scattering angle between the incident moonlight and the atmospheric particles, and the atmospheric conditions at different observatory locations.

The Rayleigh model has endured in the scientific community as a good approximation of the daytime sky polarization due to scattering of sunlight in the atmosphere \citep{Chandrasekhar60,Coulson88,liou2002introduction}. This theory has been foundational for several studies and advances, such as navigation methods based on polarized light \citep{liu2020solar} or the analysis of exoplanet atmospheres \citep{madhusudhan2012analytic, berdyugina2011polarized}. Similarly, \citet{gal2001polarization} have shown that the patterns of the degree and angle of polarization at night from the Moon are equivalent to the daylight patterns from the Sun. The Rayleigh model predicts maximum polarization at 90º from the Moon at night and two neutral points of zero polarization at the Moon and anti-Moon position. Nonetheless, changes in the relative abundances of atmospheric particles such as air molecules, aerosols and dust may result in the observation of four neutral points instead \citep{horvath2002first} due to increased multiple scattering processes \citep{berry2004polarization,wang2016analytical,Harrington17}. This strongly depends on the atmospheric conditions of the observing site.

This work is devoted to studying the polarization patterns in the moonlit sky and atmospheric scattering processes from the facilities of the European Southern Observatory (ESO) at Cerro Paranal in Chile. The scattered moonlight intensity at Paranal has been studied in detail \citep{patat2011optical,Jones19,Jones13} but not its polarization. Here we use existing analytical and empirical models of single and multiple scattering in section~\ref{sec:one} together with observations of the night sky during full Moon from Paranal in section~\ref{sec:data}, and compare them in section~\ref{sec:three} to obtain the best models and parameters that describe these patterns. In section~\ref{sec:discussion}, we discuss the implications of our results for the composition of the atmosphere and the prediction of moonlit sky polarization contamination in astronomical polarimetric observations from Paranal.

\section{Sky model}\label{sec:one}

In this section, we present the different components of the sky model used to fit our polarimetric data from Paranal (see section~\ref{sec:data}). Earth's atmosphere is composed of particles with which the moonlight of different wavelengths interacts through scattering and absorption generating the intensity and polarimetry patterns in the sky. The atmospheric transmission curve, shown in figure~1 of \citet{patat2011optical}, is characterized by the optical extinction components: Rayleigh scattering by air molecules, Mie scattering by aerosols, and molecular absorption (O$_2$, O$_3$ and H$_{2}$O). 
We closely follow the model developed for the optical intensity \citep{Jones13,Jones19}, adapting it here to optical polarimetry.

\begin{figure*}
    \centering
    \includegraphics[width=\textwidth]{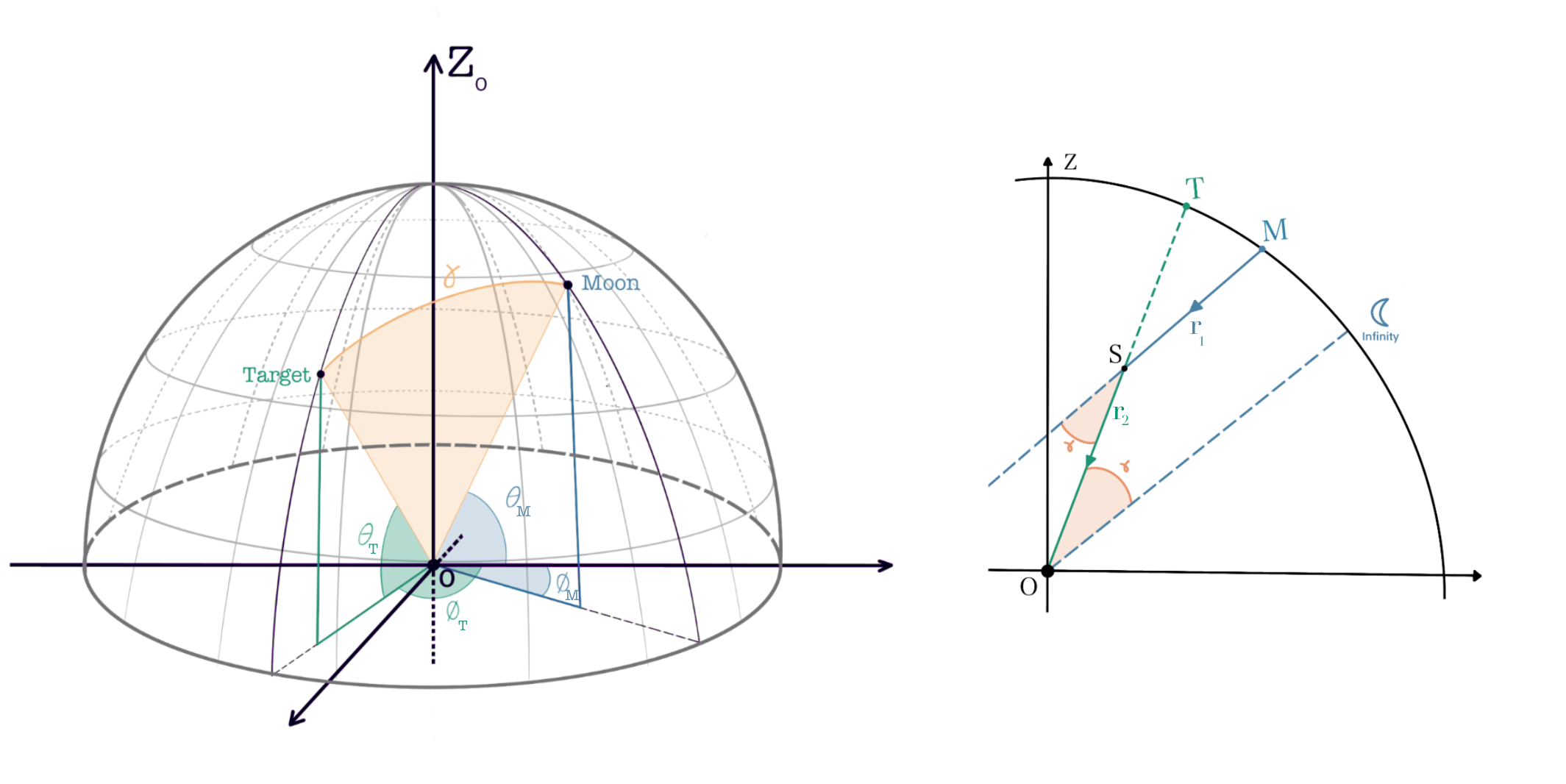}
    \caption{Left: Representation scheme for the scattering geometry in Earth’s Atmosphere. Right: Schematic description of the scattering process, considering that the moonlight enters the atmosphere in parallel beams.}
    \label{scheme}
\end{figure*}

Figure \ref{scheme} represents the Celestial Sphere, where the observer (O) is at the centre and the zenith for the observer (Z$_{\mathrm{O}}$) is in the top. $\theta$ is the altitude angle from the positive $z$-axis, and $\phi$ is the azimuth angle measured from the positive x-axis in the clockwise direction. The target that is being observed (T) and the projected Moon position on the sphere (M) form the scattering angle $\gamma$, determined as:

\begin{equation}
\small
    \cos(\gamma) = \cos(\theta_{T})\cos(\theta_{M})+\sin(\theta_{T})\sin(\theta_{M})\cos(\phi_{M}-\phi_{T})\;.
\end{equation}

The leading scattering process in the atmosphere is single scattering in which the light beam only interacts with the atmosphere once during its entire path to the observer. A light wave enters the atmosphere through M, crosses the path to a scattering point (S), in an arbitrary point on the line between the target and the observer. After interacting with the atmosphere, the photon is scattered into the line of sight towards the observer. Multiple scattering, on the other hand, represents only a few percent of the scattering \citep{Jones13} and it occurs when the light interacts several times in the atmosphere before reaching the observer.

The atmosphere contains different types of particles produced by various natural and human-induced processes. The interactions between the light and these atmospheric particles is determined by the interplay between the wavelength of light and the particle size. The latter is generally described by the size parameter $x= 2\pi a/\lambda$, where $\lambda$ is the wavelength of the incident radiation and $a$ is the typical radius of the particle. Rayleigh scattering occurs for small particles at $x<<1$ (typically air molecules including nitrogen, oxygen, and argon), whereas for larger particles (like aerosols) Mie  scattering enters into play\footnote{See e.g. figure 5 of \citet{phdthesis} for  a summary of the different wavelengths of the light and particle sizes in the atmosphere.}.

In the next sections, we review the different scattering mechanisms at play in the atmosphere. We start with some basic concepts including the incident moonlight (\S~\ref{sec:oneLight}), the Stokes formalism and scattering matrix (\S~\ref{sec:oneStokes}),  Rayleigh scattering (\S~\ref{sec:oneRay}), Mie scattering (\S~\ref{sec:oneMie}) and multiple scattering (\S~\ref{sec:oneMult}).

\subsection{Moonlight}\label{sec:oneLight}

The incident moonlight is sunlight reflected at the surface of the Moon. The solar intensity spectrum has been extensively studied \citep[e.g.][]{Colina96} while its polarization spectrum -- sometimes called the "second solar spectrum" -- shows up to a $\sim$1\% continuum polarization \citep{Stenflo97,Stenflo00,Stenflo05,Gandorfer00,Gandorfer02}. The reflection of the sunlight at the lunar surface may lead to polarization that depends on the lunar phase reaching up to 30\% at the Sun's angular distance of 90º-100º from the Moon (around first and third quarters)  and decreasing to zero at 180º or full moon \citep{Kohan62,Dollfus71,Wolff75}.
As our observations were taken during full moon, we assume in the following an unpolarized incoming moonlight spectrum as a first approximation.

\subsection{Stokes formalism and radiative transfer}\label{sec:oneStokes}

In unpolarized radiation, sometimes referred to as $natural$ light, one deals with an incoherent superposition of electromagnetic waves, meaning all orientations of the electric field vectors are equally probable. 
A convenient method to describe the polarization transformations of radiation due to scattering and other processes is through the Stokes parameters \citep{Stokes1851}. 

The Stokes vector $\vec{S}$ quantifies light polarization via four parameters (each a vector element): $I$, $Q$, $U$ and $V$. In this formalism, $I$ represents the total intensity; $Q$ and $U$ the linearly polarized intensity, more specifically $Q$ quantifies a difference in the electric field in the $x$ and $y$ reference frame (the wave propagates in the $z$-direction and usually north indicates the positive $x$-axis) and the parameter $U$ quantifies the difference between the two diagonal components (at angles of $45^{\circ}$ and $135^{\circ}$ counted from the positive $x$ axis); and $V$ refers to circularly polarized radiation.

In the absence of additional light sources, the radiation transfer equation describing changes in the radiation vector $\vec{S}$ due to its interaction with the medium can be written as \citep{Steinacker13}:

\begin{eqnarray}\label{eq:radtrans}
\frac{d\vec{S}}{dr}(r,\lambda) = &-&\kappa_{\mathrm{ext}}(r,\lambda)\delta(r)\vec{S}(r,\lambda)\\ 
&+&\kappa_{\mathrm{sca}}(r,\lambda)\delta(r)\int_{4\pi}\mathcal{M}(r,\lambda,\vec{n_1},\vec{n_2})\vec{S}(r,\lambda)d\Omega \nonumber
\end{eqnarray}

\noindent in which the left hand side is the change of the radiation over an infinitesimal part of its path $dr$, the first term of the right-hand side is the loss of energy due to extinction, and the second term represents the scattering component into the line of sight. The variable $\delta(r)$ is the mass density of the medium, $\kappa_{\mathrm{sca}}$ the scattering coefficient and $\kappa_{\mathrm{ext}}$ is the mass extinction coefficient, which includes absorption and scattering off the line of sight:  $\kappa_{\mathrm{ext}}=\kappa_{\mathrm{abs}}+\kappa_{\mathrm{sca}}$. The scattering matrix $\mathcal{M}$, or phase matrix, is the M\"uller $4\times4$ matrix  describing the changes of the Stokes vector when radiation is scattered from an initial propagation direction $\vec{n_1}$ to a new direction $\vec{n_2}$.

An incident Stokes vector $\vec{S_i}=(I_i,Q_i,U_i,V_i)$ entering at the Moon position (M) in the celestial sphere gets extincted along its path $r_1$ to the observing target (T) by a factor: $\exp(-\tau_1)$ with:
\begin{equation}
    \tau_1(\lambda) =  \kappa_{\mathrm{ext}}\int_0^{r_1}\delta(r)dr\,.
\end{equation}

In the case of single scattering, the solution to Eq.~\ref{eq:radtrans} is given by:

\begin{equation}
    \vec{S}_{\mathrm{sca}}(\lambda) = \frac{\kappa_{\mathrm{sca}}(\lambda)}{2\pi}\int_0^{r_2}\delta(r)\mathcal{M}(r,\lambda,\vec{n_1},\vec{n_2})\vec{S_i}e^{-(\tau_1+\tau_2)}dr  \label{eq:radtrans2}
\end{equation}

\noindent with $\tau_2$ indicating the extinction from S to the observer at position O along the path $r_2$. In the case of sunlit or moonlit sky observations, the only predominant contribution is that from scattering, so that $\vec{S}_{\mathrm{sca}}(\lambda)$ in Eq.~\ref{eq:radtrans2} represents the observed Stokes vector reaching the observer. 

It is important to note that in a dust mixture, both the individual coefficients and M\"uller matrices of each species, $i$, can be taken into account with their relative contributions $w_i$ in Eqs.~\ref{eq:radtrans} and \ref{eq:radtrans2} \citep{Wolf03}, as follows:

\begin{equation}\label{eq:sum-coeff}
\begin{aligned}
\kappa_{\mathrm{abs}} &=&\sum_i w_i\,\kappa_{\mathrm{abs,i}}(\lambda),\\
\kappa_{\mathrm{sca}} &=&\sum_i w_i\,\kappa_{\mathrm{sca,i}}(\lambda),\\
\kappa_{\mathrm{ext}} &=&\sum_i w_i\,\kappa_{\mathrm{ext,i}}(\lambda),\\
\end{aligned}
\end{equation}
and

\begin{equation}\label{eq:sum-muller}
    \mathcal{M}(r,\lambda,\vec{n_1},\vec{n_2}) = \frac{\sum_i w_i\,\kappa_{\mathrm{sca,i}}(\lambda)\,\mathcal{M}_i(r,\lambda,\vec{n_1},\vec{n_2})}{\sum_iw_i,\kappa_{\mathrm{sca,i}}(\lambda)}\,.
\end{equation}
\noindent In the next sections we will consider the different types of dust scattering processes contributing to Eq.~\ref{eq:radtrans2}.

\subsection{Scattering and M\"uller matrix} 
The single scattering matrix $\mathcal{M}$ that depends on the initial and final directions of light propagation, $\vec{n_1}$ and $\vec{n_2}$ respectively, can also be described via the scattering angle $\gamma$ and two $4\times4$ rotation matrices (rotation in the $Q$-$U$ plane), $\vec{R}$ \citep{liou2002introduction}:

\begin{equation}
    \mathcal{M'}(r,\lambda,\vec{n_1},\vec{n_2})=\vec{R}(\pi-\beta')\mathcal{M}(r,\lambda,\gamma)\vec{R}\,(-\beta)\;,
    \label{eq:muller-rotation}
\end{equation}

\noindent where $\beta$ and $\beta'$ are rotation angles that relate the meridian and scattering planes:

\begin{equation}\label{eq:rotangles}
\begin{aligned}
\cos\beta &=& \frac{-\cos\theta_T+\cos\theta_M\cos\gamma}{\pm\sin\gamma\sin\theta_M},\\
\cos\beta' &=& \frac{-\cos\theta_M+\cos\theta_T\cos\gamma}{\pm\sin\gamma\sin\theta_T},\\
\end{aligned}
\end{equation}
where the plus sign is used when $\pi<\phi_T-\phi_M<2\pi$ and the minus sign when $0<\phi_T-\phi_M<\pi$.

The M\"uller matrix $\mathcal{M}$ describing the scattering of light is composed of 16 independent elements, but the spherical assumptions of the particles reduce this number to four, as described in e.g. \citealt{bohren98,narasimhan2003shedding}:

\begin{equation}\label{eq:mueller}
    \begin{bmatrix}
 I\\
 Q\\
 U\\
 V
\end{bmatrix} = \frac{1}{k^2 r^{2}}\begin{bmatrix}
\mathcal{M}_{11} & \mathcal{M}_{12} & 0 & 0 \\
\mathcal{M}_{12} & \mathcal{M}_{11} & 0 & 0 \\
0 & 0 & \mathcal{M}_{33} & \mathcal{M}_{34} \\
0 & 0 & -\mathcal{M}_{34} & \mathcal{M}_{33} \\
\end{bmatrix}\begin{bmatrix}
I_i \\
Q_i \\
U_i \\
V_i
\end{bmatrix}\;,
\end{equation}

\noindent where $r$ is the distance from the observer to the scattering element and $k$ is the wave number in the medium related to the angular frequency $\omega$ and the complex index of refraction $m$ of the spherical particle: $k=\omega m/c$. To express the elements of the scattering phase matrix we use the complex scattering amplitudes $s_{1}(m,x,\gamma)$ and $s_{2}(m,x,\gamma)$, which relate to the  incoming electric field components as:
\begin{equation}
\begin{aligned}
    E_1 & \sim s_1(\gamma)\frac{\exp(ikr)}{-ikr}E_i\sin\phi \\
    E_2 &\sim s_2(\gamma)\frac{\exp(ikr)}{-ikr}E_i\cos\phi\;. 
    \end{aligned}
\end{equation}
\noindent The subscripts $1$ and $2$ correspond to two orthogonal orientations that are both perpendicular to the direction of propagation of the incoming wave.
The scattering phase matrix elements for a single sphere are then given by:

\begin{eqnarray*}
    \mathcal{M}_{11} &=& \frac{1}{2}\left [ s_{1}s_{1}^{*}+s_{2}s_{2}^{*} \right ]\\% = \frac{4\pi }{2k^{2}\sigma _{s}} (i_{1} + i_{2})\;,
    \mathcal{M}_{12} &=& \frac{1}{2}\left [ s_{2}s_{2}^{*}-s_{1}s_{1}^{*} \right ]\\%= \frac{4\pi }{2k^{2}\sigma _{s}} (i_{2} - i_{1})\;,
    \mathcal{M}_{33} &=& \frac{1}{2}\left [ s_{2}s_{1}^{*}+s_{1}s_{2}^{*} \right ]\\% = \frac{4\pi }{2k^{2}\sigma _{s}} (i_{3} + i_{4})\;,
    \mathcal{M}_{34} &=& \frac{-i}{2}\left [ s_{1}s_{2}^{*}-s_{2}s_{1}^{*} \right ].% = \frac{4\pi }{2k^{2}\sigma _{s}} (i_{4} - i_{3})\;.
\end{eqnarray*}

If the incident light is unpolarized, $\vec{S_i}=(I_i,0,0,0)$, and neglecting for now the extinction, then the scattered light has the following components: 
\begin{equation}\label{eq:scatvec}
    \vec{S} = \frac{I_i}{k^2r^2}(\mathcal{M}_{11},\mathcal{M}_{12},0,0)\,
\end{equation}
\noindent or including rotations (Eq.~\ref{eq:muller-rotation}):
\begin{equation}\label{eq:scatvec-rot}
    \vec{S} = \frac{I_i}{k^2r^2}(\mathcal{M}_{11},\mathcal{M}_{12}\cos(2\pi-2\beta'),-\mathcal{M}_{12}\sin(2\pi-2\beta'),0)\,
\end{equation}

\noindent with the intensity in either case given by:

\begin{equation}\label{eq:scatint}
I=\mathcal{M}_{11}I_i = \frac{I_i}{2} (s_2s_2^* + s_1s_1^*) = \frac{I_i}{2} (|s_1|^2 + |s_2|^2)\,,
\end{equation}

\noindent the normalized degree of polarization, $p=\sqrt{(Q^2+U^2)}/I$ or DoP, by:
\begin{equation}\label{eq:scatpol}
p=\frac{|\mathcal{M}_{12}|}{\mathcal{M}_{11}} = \frac{\left|s_2s_2^* - s_1s_1^*\right|}{s_1s_1^* + s_2s_2^*} = \frac{\left||s_2|^2 - |s_1|^2\right|}{|s_1|^2 + |s_2|^2}\,.
\end{equation}
and the polarization angle, $\chi=1/2\arctan(U/Q)$ or AoP, by:

\begin{equation}\label{eq:polang}
\chi=\beta'-\frac{\pi}{2}.
\end{equation}

For multiple particles, $i$, with different properties, the individual M\"uller matrices can be summed (see Eq.~\ref{eq:sum-muller}), and in the single scattering case the total squared moduli of the scattering amplitudes $S_1$ and $S_2$ are the sum of the amplitudes of each individual spherical particle, $s_1^i$ and $s_2^i$, taking into account its scattering coefficients \citep{liou2002introduction}:

\begin{equation}
     |S_{1,2}(\gamma)|^2 = \frac{1}{\kappa_{\mathrm{sca}}^T}\sum_i \kappa_{\mathrm{sca}}^i |s_{1,2}^i(m_i,a_i,\gamma)|^2  \; ,
\end{equation}

\noindent with $\kappa_{\mathrm{sca}}^T=\sum_i\kappa_{\mathrm{sca}}^i$.
In the case of a set of particles in the atmosphere with a distribution of sizes, the scattering amplitudes can be integrated over the number density $n(a)$ (particles per unit volume) with size parameter between $a$ and $a + da$: 

\begin{equation}\label{eq:Stot}
    |S_{1,2}(m,\gamma)|^2 = \frac{1}{\kappa_{\mathrm{sca}}^T} \int_{a_1}^{a_2} \kappa_{\mathrm{sca}}|s_{1,2}(m,a,\gamma)|^2 n(a) da  \; ,
\end{equation}

\noindent with $\kappa_{\mathrm{sca}}^T=\int \kappa_{\mathrm{sca}}(a) n(a) da$.
From these new total scattering amplitudes it is possible to obtain the intensity and the polarization degree as given by Eqs.~\ref{eq:scatint} and~\ref{eq:scatpol}.

\subsection{Rayleigh scattering}\label{sec:oneRay}

Rayleigh scattering \citep{Rayleigh1871}, also referred to Rayleigh-Gans theory, is the main explanation behind phenomena like the blue sky and the red sunrise/sunset. Originally developed for the Sun at daylight, the effects are analogous for the Moon at night. Molecules in the air (mainly nitrogen, oxygen, and argon), assumed spherical with radii $a$ much smaller than the wavelength of the incident radiation ($x<<1$), scatter preferentially light of smaller wavelengths, i.e. blue light. Such particles represent the majority of particles in the atmosphere and it is therefore the most probable type of scattering. The model assumes that the particles are identical and exist in an isotropic atmosphere.

For Rayleigh scattering, the amplitudes are given by \citep{bohren98}:
\begin{eqnarray}
    s_1^{\mathrm{ray}}(\gamma)=c_1 \;\;\;\; \mathrm{and} \;\;\;\; s_2^{\mathrm{ray}}(\gamma)=c_1\cos\gamma \\ \nonumber
    \mathrm{with} \;\;\;\; c_1=-i\left(\frac{2\pi a}{\lambda}\right)^3\frac{m^2-1}{m^2+1}
\end{eqnarray}

\noindent such that the following phase matrix can be used :

\begin{equation}
\small
    \vec{M}_{\mathrm{ray}}= \frac{c_1^2}{k^2r^2}\begin{bmatrix}
    \frac{1}{2}(1+\cos^2\gamma) & \frac{1}{2}\sin^{2}\gamma & 0 & 0\\ \nonumber
    \frac{1}{2}\sin^{2}\gamma  & \frac{1}{2}(1+\cos^2\gamma)  & 0 & 0\\ 
    0 & 0 & \cos\gamma  & 0\\ 
    0 & 0 & 0 & \cos\gamma 
    \end{bmatrix}.%\\
    %\mathrm{with} && C = \frac{16\pi^4a^6}{\lambda^4d^2}\left|\frac{m^2-1}{m^2+2}\right|^2
\label{matrixrayleigh}
\end{equation}

For unpolarized incoming light (see Eq.~\ref{eq:scatvec}), the scattered intensity for Rayleigh has a dependence on the distance ($r^{-2}$) and on the wavelength ($\lambda^{-4}$) --provided that the refractive index $m$ does not depend (or only slightly) on wavelength:
\begin{equation}
I_{\mathrm{ray}} = \frac{8\pi^4a^6}{\lambda^4r^2}\left|\frac{m^2-1}{m^2+2}\right|^2(1+\cos^2\gamma)I_i.
\end{equation}

The normalized degree of polarization or fraction of polarized light (see Eq.~\ref{eq:scatpol}) depends only on the scattering angle:

\begin{equation}
    p_{\mathrm{ray}} =\frac{\sin^{2 }\gamma}{1+\cos^{2}\gamma }\;.
\label{eq:rayleigh}
\end{equation}

\begin{figure*}
    \centering
     \includegraphics[width=0.495\textwidth]{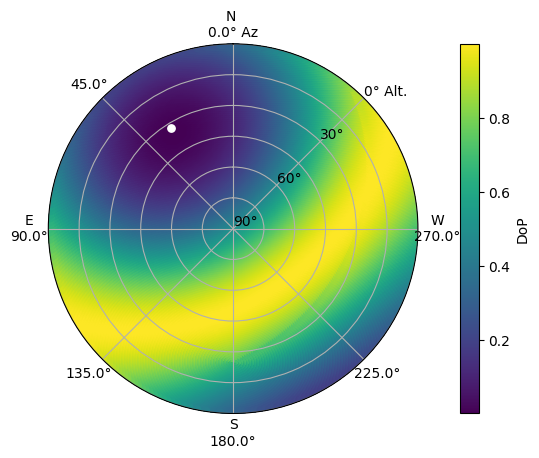}
     \includegraphics[width=0.495\textwidth]{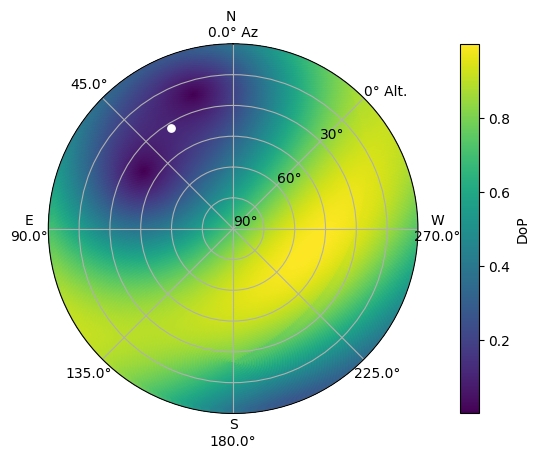}
     \includegraphics[width=0.495\textwidth]{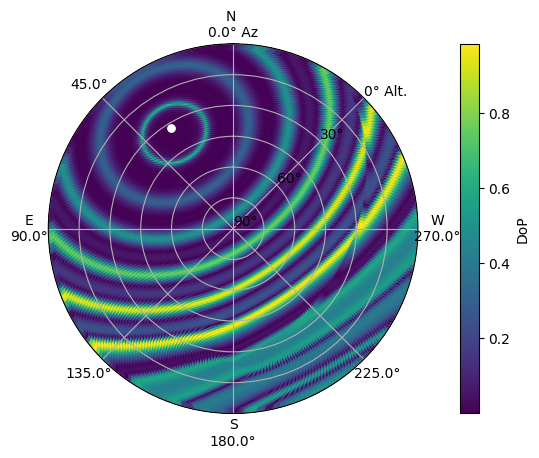}
     \includegraphics[width=0.495\textwidth]{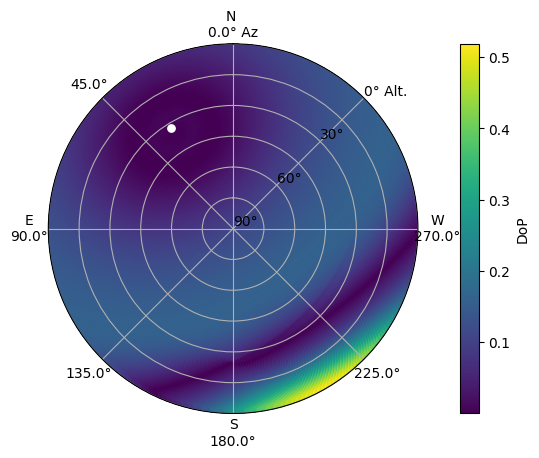}
     
    \caption{Simulation of sky polarization degree at Paranal on the 26th of January 2021, according to the single Rayleigh scattering model (\emph{upper left}),  multiple scattering model (\emph{upper right}) with $L=0.3$ or a neutral point angular distance of 67$^{\circ}$, Mie scattering model of a single size particle or radius 0.9$ \mu$m (\emph{lower left}) and of log-normal size distribution (\emph{lower right}). The polarization degree is shown in color map and the Moon position is shown as a white circle.}
    \label{fig:single-multi}
\end{figure*}

Rayleigh theory predicts therefore that in the forward and backward directions the scattered light remains unpolarized while at 90 degrees it is fully polarized; in other directions the light is partially polarized (see upper left figure~\ref{fig:single-multi}). No polarization-wavelength dependence is expected from pure Rayleigh scattering. 

\subsection{Anisotropic Rayleigh scattering}\label{sec:aniRay}

The Rayleigh scattering model presented in the previous section assumes only spherical particles, a presumption that breaks down e.g. for O$_2$, since it is diatomic and slightly anisotropic. An effective way to include anisotropic particles lies in adding the depolarization factor, $\delta$, to the Rayleigh scattering model \citep{Chandrasekhar60,Hansen1974}, such that the anisotropic amplitudes, $s_{1,2}^{\mathrm{aniray}}$, are: 

\begin{equation}
|s_{1,2}^{\mathrm{aniray}}|^2 = \Delta |s_{1,2}^{\mathrm{ray}}|^2 + |c_1|^2(1-\Delta)
\end{equation}

\noindent with $\Delta=(1-\delta)/(1+\delta/2)$, and the degree of polarization given by:

\begin{equation}
    p = \frac{\sin^{2}\gamma}{ 1+\cos^{2}\gamma +2\delta/(1-\delta)}\;.
    \label{eq:ray-depol}
\end{equation}

The depolarization factor $\delta$ is equal or smaller than 0.5. For air, experiments indicate a value of 0.03-0.04 for the depolarization factor \citep{Hansen1974}.

In the end, Rayleigh scattering is just a limiting case for infinitesimally small particles of the broader Mie framework. Larger particles in the atmosphere, such as aerosols, must be treated with Mie scattering as we do in the next section.

\subsection{Mie Scattering}\label{sec:oneMie}

Mie scattering, also known as Lorenz-Mie scattering, is a theory derived independently by Lorenz (1890) and Mie (1908), assuming that the scattering particle is a homogeneous sphere with a larger size than or equal to the wavelength of the incident ray, i.e with $x\geqslant 1$. This type of scattering is more probable for aerosols and cloud particles. As a consequence, clouds and non-absorbing aerosols in the atmosphere (like fog effects) generally appear white. The Mie scattering amplitudes are given by \citep[e.g.]{liou2002introduction}:

\begin{equation}\label{eq:s12Mie}
\begin{aligned}
    s_1^{\mathrm{mie}}(m,x,\gamma) &= \sum_n^{\infty} \frac{2n+1}{n(n+1)}(a_n\pi_n+b_n\tau_n)\\ %\nonumber
    s_2^{\mathrm{mie}}(m,x,\gamma) &= \sum_n^{\infty} \frac{2n+1}{n(n+1)}(a_n\tau_n+b_n\pi_n)\;,\\ 
    %\mathrm{with} \;\;\;\; c_1=-i\left(\frac{2\pi a}{\lambda}\right)^3\frac{m^2-1}{m^2+1}
\end{aligned}
\end{equation}

\noindent where 

\begin{equation}
\begin{aligned}
   \pi_n & = \frac{P_n(\cos\gamma)}{\sin\gamma} \\ %\nonumber
   \tau_n & =  \frac{d}{d\gamma}P_n(\cos\gamma)\;,
\end{aligned}
\end{equation}

\noindent and $P_n(\cos\gamma)$ is the Legendre polynomial of order $n$. The coefficients $a_n$ and $b_n$ are given by:

\begin{equation}
\begin{aligned}
  a_n & = \frac{\psi_n'(mx)\psi_n(x)-m\psi_n(mx)\psi'_n(x)}{\psi'_n(mx)\zeta_n(x)-m\psi_n(mx)\zeta'_n(x)} \\ %\nonumber
   b_n & = \frac{m\psi_n'(mx)\psi_n(x)-\psi_n(mx)\psi'_n(x)}{m\psi'_n(mx)\zeta_n(x)-\psi_n(mx)\zeta'_n(x)}\; ,
\end{aligned}
\end{equation}

\noindent with $\psi_n(q)$ and $\zeta_n(q)$ being quantities related to the Bessel functions $J_n(q)$ and $H_n(q)$ of the first and second kind, respectively:

\begin{equation}
\begin{aligned}
  \psi_n(q) &=  \left(\frac{\pi q}{2}\right)^2 J_{n+\frac{1}{2}}(q)\\ 
  \zeta_n(q) &= \left(\frac{\pi q}{2}\right)^2 H_{n+\frac{1}{2}}(q).
\end{aligned}
\end{equation}

The polarization degree for Mie scattering can be calculated using the scattering amplitudes (Eq.~\ref{eq:s12Mie}) in Eq.~\ref{eq:scatpol}. The resulting pattern depends not only on the scattering angle but also on the size parameter, i.e. the relation between the radius of the spherical particle and the wavelength.  

However, in reality the aerosols of the atmosphere come in several types and follow a log-normal size distribution \citep{Williams12}. The six species of tropospheric and stratospheric aerosols considered in \citet{Jones19} are shown in Table~\ref{tab:aerosols}. With a log-normal size distribution of the form:

\begin{equation}\label{eq:sizedist}
\frac{dN(r)}{d\log r} = \frac{1}{\sqrt{2\pi}}\frac{n_i}{\log s_i} \exp\left({-\frac{\log^2(r/a_i)}{2\log^2s_i}}\right)\;,
\end{equation}

\noindent where the characteristic number density $n_i$, mean radius $a_i$ and the size distribution $s_i$ per species $i$ are also shown in Table~\ref{tab:aerosols}, we can use Eq.~\ref{eq:Stot} to model their scattered amplitude as the integral of each individual particle $s_{1,2}^i$ from Eq.~\ref{eq:s12Mie}:

\begin{equation}\label{eq:sizedist}
|S_{1,2}^{i}|^2 = \frac{1}{\kappa_{\mathrm{sca}}^i}\int \kappa_{\mathrm{sca}}^i|s_{1,2}^i|^2 \frac{2\pi r}{\lambda}dN(r)\, d\log r\;,
\end{equation}

 Each species may also have a different refractive index $m_i$. An example of Mie scattering for large particles of both, identical size ($a=0.9\mu$m -- tropospheric coarse) and following a size distribution of Eq.~\ref{eq:sizedist}, is shown in the bottom left and right of Figure~\ref{fig:single-multi}. 

The total Mie single scattering phase amplitudes from all integrated species are the sum of its six constituents (the relative weights already being determined by their density):

\begin{equation}\label{eq:S12_mie}
|S_{1,2}^{\mathrm{mie}}(\gamma,\lambda)|^2 = \frac{1}{\kappa_{\mathrm{sca}}^{\mathrm{mie}}}\sum_i \kappa_{\mathrm{sca}}^i|S_{1,2}^i(m_i,\gamma,\lambda)|^2\;,
\end{equation}

Finally, in order to consider all single scattering (SS) events, including Rayleigh and Mie scattering, the  total amplitudes are given by:

\begin{equation}\label{eq:SS}
|S_{1,2}^{\mathrm{SS}}|^2 =\frac{ w_{\mathrm{ray}}\kappa_{\mathrm{sca}}^{\mathrm{ray}}|S_{1,2}^{\mathrm{ray}}|^2+w_{\mathrm{mie}}\kappa_{\mathrm{sca}}^{\mathrm{mie}}|S_{1,2}^{\mathrm{mie}}|^2}{w_{\mathrm{ray}}\kappa_{\mathrm{sca}}^{\mathrm{ray}}+w_{\mathrm{mie}}\kappa_{\mathrm{sca}}^{\mathrm{mie}}}\;,
\end{equation}

\noindent which can be in turn used to calculate the intensity and polarization degree via Eqs.~\ref{eq:scatint} and~\ref{eq:scatpol}. The relative weights of each component sum to one, i.e. $w_{\mathrm{ray}}+w_{\mathrm{mie}}=1$.

For the single size Mie scattering we use the {\sc miepython} software \citep{miepython}\footnote{\url{https://miepython.readthedocs.io/}} and for the log-normal distributed Mie scattering we implement a {\sc python} version of the {\sc IDL} code developed by G. Thomas\footnote{\url{http://www.atm.ox.ac.uk/code/mie/mie_lognormal.html}}.

\begin{table}
%\begin{threeparttable}
{\centering
\small
\caption{Aerosol modes from \citet{Williams12}}
\label{tab:aerosols}
\begin{tabular}{l||l|l|l|l|}
Type & $n$ (cm$^{-3}$) & $a$ ($\mu$m) & $\log s$ & $w_i$\\
\hline
Trop. nucleation & 3.20$\times10^3$ & 0.010 & 0.161 & 0.4\\
Trop. accumulation & 2.90$\times10^3$ & 0.058 & 0.217 & 0.18 \\
Trop. coarse & 3.00$\times10^{-1}$ & 0.900 & 0.380 & 0.02\\
Stratospheric & 4.49$\times10^0$ & 0.217 & 0.248 & 0.4\\
Trop. dust acc. & 1.14$\times10^3$ & 0.019 & 0.770 & 0.0\\
Trop. dust coarse & 1.87$\times10^{-1}$ & 1.080 & 0.438 & 0.0 \\
\hline
\end{tabular}}
%\begin{tablenotes}
\small
Trop. refers to tropospheric, acc. to accumulation. See Eq.~\ref{eq:sizedist} for the parameters; $w_i$ are the relative aerosol fractions from \citet{Jones13}.
%\item $^{\dagger}$ Values for \naid. The exact cuts and number of SNe/spectra changes according to line species. 
%\end{tablenotes}
%\end{threeparttable}

\end{table}

\subsection{Multiple scattering}\label{sec:oneMult}

The scattering in our atmosphere as seen from Paranal has only a few percent contribution from multiple scattering (MS) effects, as shown in \citet{Jones13}. Doing a full physical treatment of multiple scattering is very difficult. Here we experiment with two approaches: 1) a simple phenomenological model of the polarization pattern originating from multiple scattering; and 2) the assumption that an additional component of light from multiple scattering is non-polarized.

\subsubsection{Phenomenological multiple scattering polarization}\label{sec:berry}
Multiple scattering in the atmosphere has been supported by observations that often show the existence of more neutral polarization points \citep{Brewster1847,horvath2002first} than what the single Rayleigh scattering predicts. The four predicted neutral points are located close to the Moon and anti-Moon positions (but not coincident with them), and they can be modeled with radiative transfer assuming multiple-scattering effects \citep{Chandrasekhar50,Chandrasekhar54}. The Berry singularity theory \citep{berry2004polarization} provides a simple alternative phenomenological description of the sky polarization pattern with four neutral points that is consistent both with observations and with radiative transfer calculations. Here, the polarization can be represented by a complex Stokes parameter:

\begin{equation}
    \omega(\xi) = \left< (E_{x}+iE_{y})^{2}\right> = \left|\omega(\xi) \right|e^{2i\gamma (\xi )}\;.
\end{equation}

Then the degree of polarization will be given by $ p_{\mathrm{MS}}=\left| \omega(\xi)\right|$ and the polarization direction or the polarization angle is given by $\gamma (\xi )$. With this description the four neutral points known as Brewster and Babinet points (close the Moon), and Arago and second Brewster (close to the anti-Moon) are described by the points $\xi _{+}$, $\xi _{-}$, $\frac{-1}{ \xi _{+}^{*}}$ and $-\frac{1}{ \xi _{-}^{*}}$, respectively. To guarantee that the degree of polarization is antipodally invariant, it is imposed that $| \omega(\frac{-1}{\xi ^{*}})|=| \omega(\xi)|$, meaning the function $\omega(\xi)$ becomes:

\begin{equation}
\small
    \omega(\xi )= 4\frac{(\xi  - \xi _{+} )(\xi -\xi _{-})\left ( \xi + \frac{1}{\xi _{+}^{*}} \right )\left ( \xi +\frac{1}{\xi _{-}^{*}} \right )}{(1+r^{2})^{2}\left | \xi _{+}+\frac{1}{\xi _{+}^{*}} \right |\left | \xi _{-}+\frac{1}{\xi _{-}^{*}} \right |}\;,\label{eq:mult}
\end{equation}

\noindent with 
\begin{equation}
\small
\begin{aligned}
   \xi_+ & =  \frac{x'_M + L \cos(\phi_M)}{1-x'_M L\cos(\phi_M) } + i \frac{y'_M + L \cos(\phi_M)}{1- y'_M L \cos(\phi_M)}\,, \\
   \xi_- & =  \frac{x'_M - L \cos(\phi_M)}{1+ x'_M L \cos(\phi_M)} + i \frac{y'_M - L \cos(\phi_M)}{1+y'_M L \cos(\phi_M)}\,,
\end{aligned}
\end{equation}

\noindent where $x'_{M}=x_M/(1-z_M)$ and $y'_{M}=y_M/(1-z_M)$. $x_M$, $y_M$ and $z_M$ are the coordinates of the Moon, $\phi _{M}$ is the azimuth of the Moon and $L$ is a parameter that is used to describe the degeneration of the two singularities of $w(\xi)$. The angular separation between the two singularities is given by $\delta=4\arctan(L)$ \citep{wang2016analytical}. We use a normalization parameter $A$ that multiplies $w(\xi)$ and sets the maximum degree of polarization. In this phenomenological model the sky pattern for the degree of polarization results in the upper right figure \ref{fig:single-multi}. However, it does not provide the scattering amplitudes and it is then difficult to combine with other mechanisms. Here we fit it individually to our data to see if it is the leading process in the atmosphere for our observations.

\subsubsection{Multiple scattering intensity}\label{sec:mult}

For the intensity contribution from multiple scattering, we use the results of \citet{Jones13}, which give a correction factor $f(\lambda,\gamma)$ for the single scattering flux to account for the multiple scattering: $I^{\mathrm{tot}} = I^{\mathrm{SS}}f(\lambda,\gamma)$. 

If we combine this intensity with the polarization pattern predicted by the multiple-scattering phenomenological model, we can posit the following scattering amplitudes that ensure both the polarization of \citet{berry2004polarization} and the intensity of \citet{Jones13}:

\begin{equation}\label{eq:S12_MS}
\begin{aligned}
   |S_1^{\mathrm{MS}}|^2 = \frac{1}{2}\left[|S_1^{\mathrm{SS}}|^2+|S_2^{\mathrm{SS}}|^2\right]\left(f(\lambda,\gamma)-1\right)\,[1-\omega(\xi)]\,\,\\
   |S_2^{\mathrm{MS}}|^2 = \frac{1}{2}\left[|S_1^{\mathrm{SS}}|^2+|S_2^{\mathrm{SS}}|^2\right]\left(f(\lambda,\gamma)-1\right)\,[1+\omega(\xi)]\,;
\end{aligned}
\end{equation}

\noindent where $\omega(\xi)$ is given by Eq.~\ref{eq:mult}. 

In general, higher-order scattering leads to increased depolarization because of the growing randomness in the direction of scattering of the light. If $\omega=0$, multiple scattering is unpolarized, which means no preferential orientation of the electric field, and the scattering amplitudes would be equal: $S_1^{\mathrm{MS}}=S_2^{\mathrm{MS}}$. The polarization contribution from multiple scattering will thus be null, $p^{\mathrm{MS}}=0$ but not the intensity, $I^{\mathrm{MS}}=|S_1^{\mathrm{MS}}|^2$.

\subsection{Total scattering}

When we consider all scattering processes, single Rayleigh and Mie scattering, as well as multiple scattering, and also the extinction factors, we can re-write the total polarization degree and intensity as:

\begin{eqnarray} 
p_{\mathrm{tot}} &=&  \frac{\left||S_2^{\mathrm{tot}}|^2 -|S_1^{\mathrm{tot}}|^2\right|}{|S_2^{\mathrm{tot}}|^2 +|S_1^{\mathrm{tot}}|^2} \\
I_{\mathrm{tot}} &=& \frac{I}{2}\left(|S_1^{\mathrm{tot}}|^2 +|S_2^{\mathrm{tot}}|^2\right)e^{-\kappa_{\mathrm{ext}}}
\end{eqnarray}

\noindent with

\begin{equation}\label{eq:S12_tot}
|S_{1,2}^{\mathrm{tot}}|^2 = 
|S_{1,2}^{\mathrm{SS}}|^2+w_{\mathrm{MS}}|S_{1,2}^{\mathrm{MS}}|^2\,,
\end{equation}

\noindent and $|S_{1,2}^{\mathrm{SS}}|^2$ given by Eq~\ref{eq:SS} and the weights of each process, $w$, are free parameters. For simplification in the fits performed in this study, we concentrate on the dependence on wavelength, particle size and refractive index and assume that the integrations of path are mostly absorbed by a general constant $I$ left free in the fit.

\section{Data Acquisition}\label{sec:data}

In this section, we present the strategy to observe the moonlit sky with the imaging polarimetric mode (IPOL) of the Focal Reducer and low dispersion Spectrograph (FORS2) at the Very Large Telescope (VLT) in Paranal, as well as the data acquisition and reductions we performed.

\subsection{Observational strategy}
To test the different atmospheric models presented in the previous section, we took polarimetric observations in $BVRI$ filters of fourteen blank fields during the full-moon nights of $25^{th}$ and $26^{th}$ of January 2021 using FORS2 at VLT through program 106.21L9. Blank fields refer to fields of view devoid of bright sources and are generally used as flat fields for calibrations. Since the polarization pattern for different models changes mostly with scattering angle (see Figure~\ref{fig:single-multi}), we tried to cover a wide range of distances from the Moon including also targets close to the Moon to test for neutral points.

\begin{figure}[h]
\centering
\includegraphics[trim={-0.5cm 0.7cm -0.5cm 0.7cm},width=.5\textwidth]{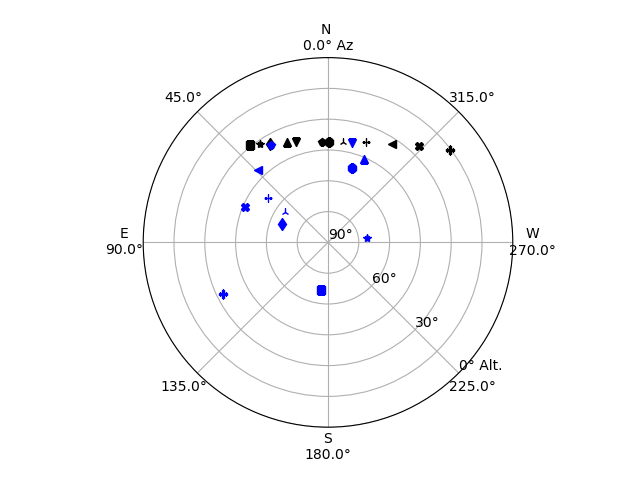}
\caption{Summary of observations as seen from an observer at Cerro Paranal during two consecutive nights. The zenith is at the center and the blank fields in the sky (our targets) at time of observation are shown with blue markers. The corresponding Moon position at the time of observation of each target is shown in black with the same symbol shape.}
\label{observations}
\end{figure}

Our observations are summarized in Figure~\ref{observations}, where the field locations are shown in blue, and the Moon positions during the night are represented in black with matching symbols with the field at the time of its observation. We aim to measure the illuminated sky and therefore the exposure times were calculated based on the number of counts ($>$30k on average per frame), so that fields farther from the Moon had longer integration times. The average exposure times per HWP angle of all our observed fields are 198, 94, 63, and 63 seconds for $B$, $V$, $R$ and $I$, respectively.
 
 On the first night of observations, the weather started with thin cirrus and windy, it evolved to higher seeing values until getting cloudy with thick cirrus and temporarily closing the dome. On the second day of observations, the observing conditions were optimal. Measurements were taken until close to twilight, when the Sun started influencing the polarization pattern. Due to different observation conditions, the data were divided into three different groups: a) clear data, b) cloudy data and c) twilight data. The median seeing of all our observations was 0.87''.

 We note that these observations can also help to measure the stability of the FORS2 spatial instrumental corrections found in  \citet{Gonzalez-Gaitan20}, as will be presented elsewhere.

\subsection{Polarimetric measurements}

The FORS2 instrument is a dual-beam polarimeter, in which one measures simultaneously the intensity of the two perpendicular polarization components of the incident light beam. The beam is divided in two by a Wollaston prism (WP) whereas the rotatable half-wave plate (HWP), placed before the WP, allows measurement at different position angles $\theta_{i}$. With the HWP at a position $i$, two intensities designed as ordinary $f_{O,i}$ and extraordinary $f_{E,i}$ beams, are derived. We took here four HWP angles for every field:  $ 0^{\circ}$, $22.5$, $45^{\circ}$ and $67.5^{\circ}$.

Since the fields are evenly illuminated and the polarization is not expected to change substantially in the 7' field-of-view, all pixels can be used to obtain the polarization of the field. As in \citet{Gonzalez-Gaitan20}, we bin the fluxes in $30\times30$ pixels in $x$ and $y$ by taking the median and the standard deviation as the characteristic value and uncertainty of each box.  

The Stokes parameters can then be computed from these set of binned beams, either through the double-difference or the double-ratio methods \citep{Bagnulo09}. Note that we assume only linear polarization and therefore $V = 0$. The normalized binned parameters are designated as $q_b$ and $u_b$, and correspond to the parameter divided by the total intensity: $q_b \equiv Q_b/I_b$ and $u_b \equiv U_b/I_b$, respectively. In the double-difference method:

\begin{equation} 
\begin{aligned}
    q_b(x,y) &= \frac{2}{N}\sum_{i=0}^{N-1}\frac{f_{O,i}-f_{E,i}}{f_{O,i}+f_{E,i}} \cos\left ( \frac{\pi }{2}i \right ) \\
    u_b(x,y) &= \frac{2}{N}\sum_{i=0}^{N-1}\frac{f_{O,i}-f_{E,i}}{f_{O,i}+f_{E,i}} \sin\left ( \frac{\pi }{2}i \right )\;,
\end{aligned}
\end{equation}

\noindent where $N=4$ is the total number of HWP position angles. 
To obtain a final value for the entire field, we take the median and standard deviation of all binned $q_b,u_b$ values: 

\begin{equation}
\begin{aligned}\label{eq:median}
q,\delta_q &= \mathrm{MED}(q_b), \mathrm{SD}(q_b), \\
u,\delta_u &= \mathrm{MED}(u_b), \mathrm{SD}(u_b)
\end{aligned}
\end{equation}

There is a spurious spatial instrumental polarization pattern that is axisymmetric in both $q$ and $u$ \citep{Gonzalez-Gaitan20} and thus in principle removed with the median. However, we cross-check that the values are consistent when doing first an instrumental field correction to the binned $q_b,u_b$ values before taking the median (Eq.~\ref{eq:median}).

The polarization degree and polarization angle are then given by:

\begin{equation}
\begin{aligned}
p & = \sqrt{q^2+u^2},\\
\psi &= \frac{1}{2}\arctan\frac{u}{q}
\end{aligned}
\end{equation}

The errors associated with the degree and angle of polarization can be obtained by propagating the uncertainties as in the following relation:

\begin{equation}
\begin{aligned}
\delta p &= \frac{\sqrt{(q \cdot \delta q) ^2 + (u \cdot \delta u) ^2}}{p}\,, \\
    \delta \psi &= \frac{1}{2} \frac{\sqrt{(q \cdot \delta u)^2 + (u \cdot \delta q)^2}}{q^2 + u^2}\; ,
\end{aligned}
\end{equation}

\noindent where $\delta q$ and $\delta u$ are the uncertainties associated to $q$ and $u$.  The full list of obtained quantities are presented in appendix~\ref{ap:data}.

\section{Results}\label{sec:three}

In this section, we compare the theoretical scattering predictions with our FORS2-IPOL observations. The observations were done during two nights near full Moon and the Moon's phase variation during those two nights of observation is assumed to be irrelevant in the models. We used not only the polarization degree observations but also the intensity measurements that will help disentangle the various models. The polarization angle is largely independent of the atmospheric models (see Eq.~\ref{eq:polang}) and is shown in Appendix~\ref{ap:polang}. We focus here on results obtained from data acquired in clear sky conditions, and refer the reader to Appendix~\ref{sec:atm} for the atmospheric corrections and Appendix~\ref{sec:Apthree} for twilight data.  

Our fits are Levenberg-Marquardt minimizations through the {\sc lmfit} package\footnote{\url{https://zenodo.org/records/15014437}} \citep{newville2016lmfit} followed by posterior sampling with a Markov Chain Monte Carlo (MCMC, see Appendix~\ref{ap:mcmc}) with the {\sc emcee} package \citep{emcee13}. To compare models with a different number of free parameters, we make use of the Bayesian Information Criterion (BIC), a diagnostic of the quality of the fit that penalizes a higher number of free parameters in the model. Lower values indicate model preference, especially when the difference between two models, $\Delta$BIC, is larger than 10.

\subsection{Rayleigh scattering}
Here we fit the Rayleigh scattering model first to the polarization degree, then the intensity and finally to both, polarization and intensity, simultaneously.

\begin{figure*}[h]
    \centering
    \includegraphics[width=.465\textwidth]{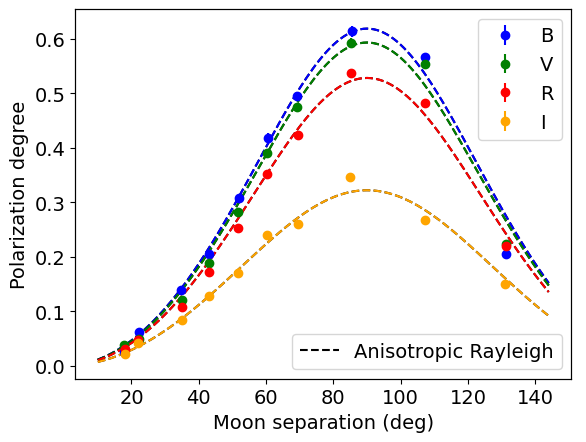}
    \includegraphics[width=.465\textwidth]{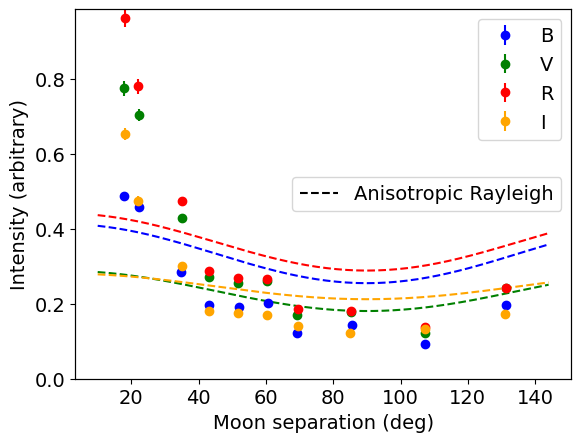}
    \caption{Polarization degree (left) and intensity in arbitrary units (right) versus scattering angle from the Moon. Points are observations in different bands (uncertainties are smaller than the points: $\sim$0.001-0.01 for polarization, and $\sim$0.001-0.02 for intensity) and lines are individual fits to each band for anisotropic Rayleigh scattering.}
    \label{plot_rayleigh}
\end{figure*}

\begin{enumerate}[i]
\item \textbf{Polarization fits}: 
 As can be seen in the left panel of Figure~\ref{plot_rayleigh}, the polarization degree follows a scattering angle pattern very similar to the Rayleigh phase function peaking near 90$^{\circ}$ and close to zero at forward and backward directions (see Eq.~\ref{eq:rayleigh}). However, pure single Rayleigh scattering predicts 100\% peak polarization at all wavelengths, which is not seen here. In fact, in order to fit a Rayleigh model to the data, we need to assume anisotropic grains that strongly depolarize the light with a wavelength-dependent factor $\delta(\lambda)$, as in Eq.~\ref{eq:ray-depol}, the only free parameter in each fit. The fitted models for each band are represented with dashed lines in the left panel and the parameters and BIC values are shown in Table~\ref{tab:polmodels}. Clearly, the depolarization is stronger as we go to longer wavelengths increasing from $\delta = 0.03$ to 0.12 following a power-law (see Appendix~\ref{ap:powlaw}). The bluer wavelengths agree with measured atmospheric values (e.g. 0.02 for H$_2$, 0.03 for air and N$_2$, 0.06 for O$_2$ and 0.09 for CO$_2$; see e.g. \citealt{Penndorf57}), whereas the reddest bands show clear signals of increased depolarization, probably from non-spherical particles (such as ice crystals, snow flakes or dust particles) or from enhanced multiple scattering \citep{Sassen91}. Other effects beyond pure Rayleigh scattering become more dominant with wavelength as indicated by the growing depolarization. The decreasing BIC values with wavelength confirm that models with deviations from pure Rayleigh due to depolarization are preferred.

\item \textbf{Intensity fits}:
As can be seen in the right panel of Figure~\ref{plot_rayleigh}, the intensity grows strongly at small scattering angles, a pattern that is more reminiscent of the Mie scattering phase function and inconsistent with Rayleigh scattering. Doing individual fits to each band's intensity, we obtain the dashed lines shown in the right panel, which clearly under-predict the intensity at small angles and result in high BIC values as shown in Table~\ref{tab:intmodels}. We note that the intensity is normalized by a global factor that preserves its wavelength and scattering angle dependence. The Rayleigh model thus has an additional free normalization parameter for the intensity for a total of two parameters ($\delta$ and $I$). We note that the depolarization factor does not have an effect on the intensity. 

\item \textbf{Polarization and intensity fits}:
A simultaneous fit to both, polarization and intensity data, for each band results in the dotted lines shown in Figures~\ref{plot_polint1} and~\ref{plot_polint2}. These are isotropic Rayleigh fits with no depolarization factor and thus peaking at 100\% polarization degree. 
\end{enumerate}
We conclude that Rayleigh scattering alone is unable to reproduce the observations.    

\begin{figure*}[h]
    \centering
    \includegraphics[width=.465\textwidth]{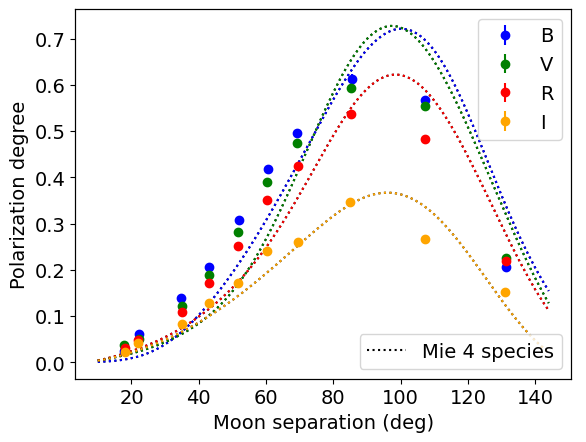}
    \includegraphics[width=.465\textwidth]{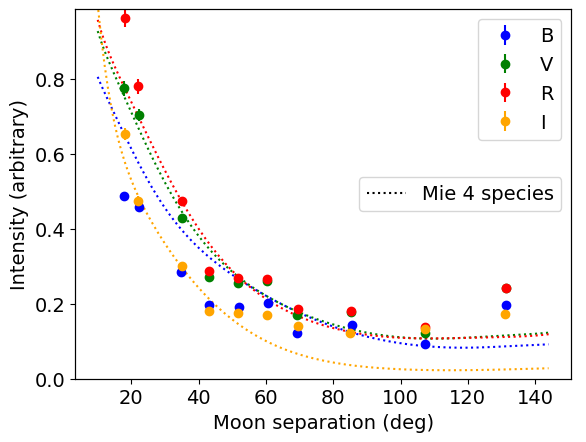}
    \caption{Polarization degree (left) and normalized intensity (right) versus scattering angle from the Moon. Points are observations in different bands (uncertainties are smaller than the points) and lines are individual fits to each band for Mie scattering with four species and a single refractive index.}
    \label{plot_mie}
\end{figure*}

\begin{table}[h]
\begin{threeparttable}
    \centering
    \caption{Polarization fit results}
    \small
    \renewcommand{\arraystretch}{1.2}
    \begin{tabular}{l||c|c|c|c}
                   & \textbf{B}
                   & \textbf{V}
                   & \textbf{R}
                   & \textbf{I} \\
    \hline
     \multicolumn{5}{c}{\textbf{Anisotropic Rayleigh}} \\
    
    $\delta$$^{\dag}$  &  0.035(02)  &  0.039(02) & 0.051(02) & 0.111(30) \\
    BIC &  $-71.5$  & $-74.8$  & $-80.8$ & $-84.5$ \\
    \hdashline
    \multicolumn{5}{c}{\textbf{Mie (4 species)}} \\
    $m$ &  $1.80(09)$ & $1.43(05)$ & $1.44(03)$ & $1.58(01)$ \\
    BIC &  $-46.9$  & $-47.1$  & $-52.7$ & $-66.7$ \\
    \hdashline
    \multicolumn{5}{c}{\textbf{Multiple scattering}} \\
    $L$ & $0.00(09)$ & $0.00(06)$ & $0.00(08)$ & $0.08(12)$\\
    BIC & $-54.8$ & $-55.7$ & $-59.7$ & $-74.9$ \\
    \hdashline
    \multicolumn{5}{c}{\textbf{Rayleigh \& Mie}} \\
    $m$ &  $1.95(16)$ & $2.06(06)$ & $2.49(11)$ & $3.29(01)$ \\
    $R/M$ & 12.97(1.24) & 12.71(0.68) & 6.39(0.70) & 0.04(2.29)\\%)447 & 73 & 7 & 4 \\ 
    BIC &  $-44.6$  & $-48.5$  & $-55.3$ & $-67.8$ \\
    \hdashline
    \multicolumn{5}{c}{\textbf{Rayleigh \& Mie \& unpol-MS}} \\
    $m$ &  $1.65(52)$ & $1.48(10)$ & $1.64(06)$ & $1.46(44)$ \\
    $R/M$ & 38.04(81) & 12.79(33) & 8.81(17) & 3.20(90) \\
    MS & $0.76(10)$ & $1.15(10)$ & $1.38(08)$ & $3.74(82)$\\
    BIC &  $-68.1$  & $-73.9$  & $-85.7$ & $-83.2$ \\
    \end{tabular}
    \label{tab:polmodels}
    \tablefoot{Summary of various models (Rayleigh, Mie, multiple scattering, combined Rayleigh and Mie, combined Rayleigh, Mie and unpolarized MS) fits to individual band polarization data. BIC values and some relevant fit parameters are shown: the depolarization factor $\delta$ for anisotropic Rayleigh scattering, the refractive index $m$ for Mie scattering, the singularity separation $L$ for polarized multiple scattering, the ratio of Rayleigh to Mie weights, $R/M = w_R/w_M$, the MS weight MS=$w_{MS}$.}
\begin{tablenotes}
\small
%\item $^{\dag}$ T
%\item $^{\ast}$ The fit uncertainties in these parameters are orders of magnitude larger than the values and are not quoted.
\item $^{\dag}$ The $\delta$ parameter is only present for anisotropic Rayleigh scattering (see section~\ref{sec:aniRay}).
\end{tablenotes}
\end{threeparttable}
\end{table}

\subsection{Mie scattering}

We now show the results of pure Mie scattering fitted to the data: polarization, intensity and joint polarization-intensity fits. We concentrate on Mie scattering with four aerosol species (the first four in Table~\ref{tab:aerosols}) and fix their respective fractions to the values obtained by \citet{Jones13,Jones19}. We assume that all species have the same refractive index (without the imaginary component, i.e. the absorption coefficient). These constraints will be relaxed in section~\ref{sec:multiwave}.

\begin{enumerate}[i]
%\subsubsection{Polarization fits}
\item \textbf{Polarization fits}:
 For Mie scattering we obtain the polarization fits shown in the left panel of Figure~\ref{plot_mie}. The fits are much worse than for Rayleigh scattering with a peak in polarization degree at higher angles than the observed $\sim90^{\circ}$. The worse fits are also seen in much higher BIC values ($\Delta$BIC$>15$). The obtained refractive index ranges between $1.4-1.5$, in agreement with measured values \citep{Elterman66,Yue94}; however in $B$, the refractive index is too high, $m=1.80\pm0.09$, indicating that Mie is even less compatible at these wavelengths. This is expected as aerosol particles responsible for Mie scattering are larger with higher cross sections at longer wavelengths of light. The fits are indeed also better at increasing wavelengths according to the BIC values. 

\item \textbf{Intensity fits}:
Regarding the intensity fits, the Mie scattering model is capable of reproducing the sharp increase at small scattering angles providing much better fits than pure Rayleigh scattering (also seen in $\Delta$BIC$<-10$ of Table~\ref{tab:intmodels}). However, at larger scattering angles, the Mie models underestimate the observed intensities. The fitted refractive indices are rather large ($m>1.8$), except for the $I$-band, showing that Mie scattering alone cannot reproduce the intensity data. The $I$-band wavelength is more compatible with observed refractive indices ($1.29\pm0.09$) pointing again to a predominance of Mie at longer wavelengths. 

\item \textbf{Polarization and intensity fits}:
In Figures~\ref{plot_polint1} and~\ref{plot_polint2}, and Table~\ref{tab:polintmodels} we confirm the fits with only polarization and the ones with only intensity data: although Mie scattering provides overall better fits than Rayleigh scattering, a combination of both processes probably reconciles the shortcomings of both. 
\end{enumerate}

\subsection{Multiple scattering}
\begin{figure}[h]
    \centering
    \includegraphics[width=.495\textwidth]{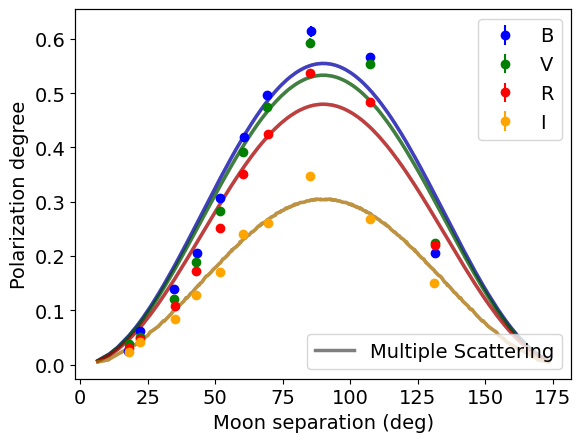}
    \caption{Polarization degree versus scattering angle from the Moon. Points are observations in different bands (uncertainties are smaller than the points) and lines are individual fits to each band for the phenomenological multiple scattering model.}
    \label{plot_poldeg_ms}
\end{figure}

\begin{figure*}[h]
    \centering
    \includegraphics[width=.465\textwidth]{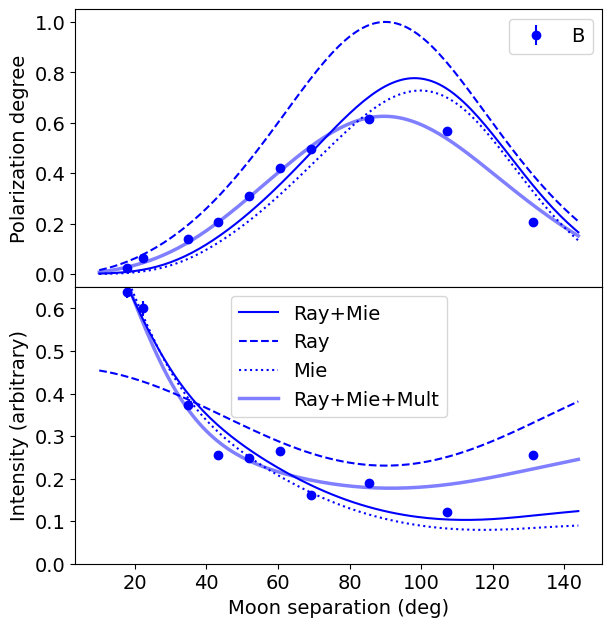}
    \includegraphics[width=.465\textwidth]{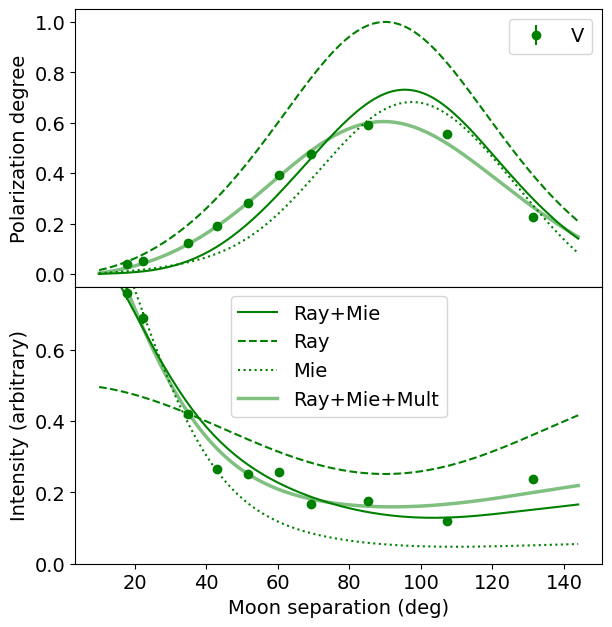}
    \caption{Polarization degree (upper) and normalized intensity (lower) versus scattering angle from the Moon for $B$ (left) and $V$ (right) bands. Points are observations in different bands and lines are polarization-intensity fits to each band. Error bars are smaller than the data points ($\sim$0.001-0.01 for polarization, and $\sim$0.001-0.02 for intensity).}
    \label{plot_polint1}
\end{figure*}

\begin{figure*}[h]
    \centering
    \includegraphics[width=.465\textwidth]{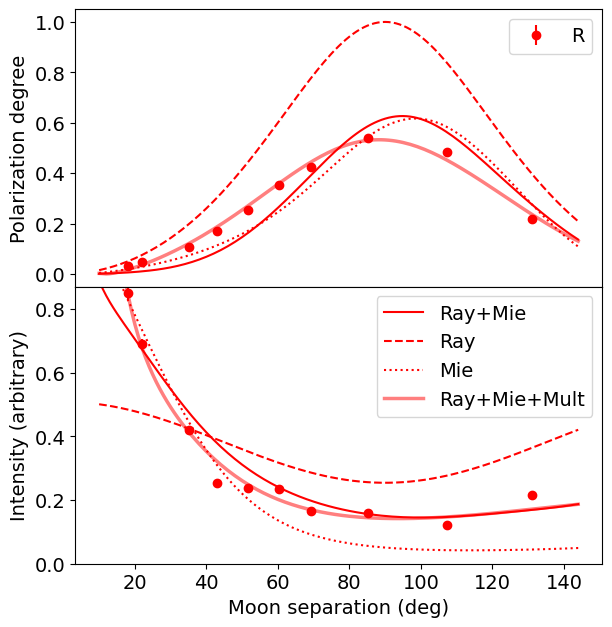}
    \includegraphics[width=.465\textwidth]{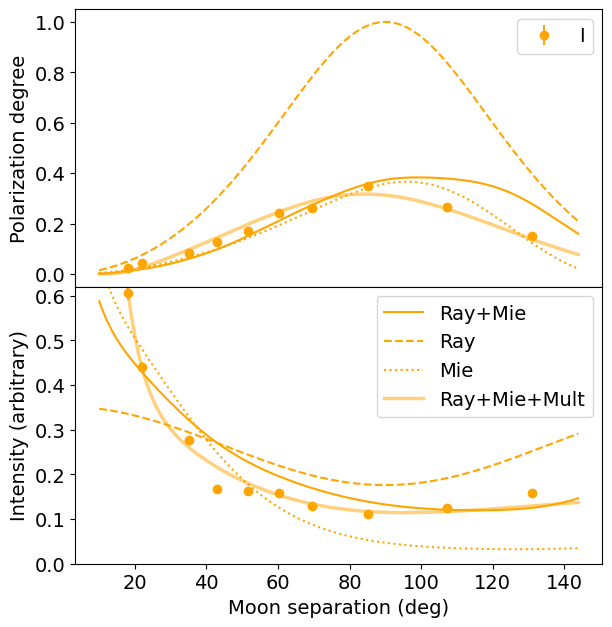}
    \caption{Same as Figure~\ref{plot_polint1} for $R$ (left) and $I$ (right) bands.}
    \label{plot_polint2}
\end{figure*}

We also fit the phenomenological multiple scattering model of Eq.~\ref{eq:mult} to our observed polarization data as shown in Figure~\ref{plot_poldeg_ms}. In this case the deviation from pure Rayleigh is given by the $L$ parameter, a measure of the distance between the two neutral points (see \S~\ref{sec:berry}). We recover single scattering, i.e. $L=0$, for $BVR$ bands. For the $I$ band, however, the value is larger than zero, although within the uncertainties ($L=0.08\pm0.12$). Such a value of $L=0.08$ corresponds to 18$^{\circ}$ angular distance between neutral points, which is slightly lower than what is found at other observing sites \citep[22$^{\circ}$-40$^{\circ}$,][]{horvath2002first}. Non-zero $L$ values for the $I$-band confirm that multiple scattering is more relevant at longer wavelengths because of the presence of larger atmospheric particles (like aerosols). Similarly as with anisotropic Rayleigh and Mie scattering, the BIC values also reveal better fits at longer wavelengths. We note, however, that to constrain this model better, more data near the neutral points is required. 

Since our Paranal fits result in insignificant $L$ values, we assume here that multiple scattering does not induce significant polarization but rather only depolarizes the light. As such, hereafter, multiple scattering is added as an intensity component without polarization signature ($\omega=0$ in Eq.~\ref{eq:S12_MS}).

\begin{table}[h]
    \centering
    \caption{Intensity fit results}
    \small
    \renewcommand{\arraystretch}{1.2}
    \begin{tabular}{l||c|c|c|c}
                   & \textbf{B}
                   & \textbf{V}
                   & \textbf{R}
                   & \textbf{I} \\
    \hline
     \multicolumn{5}{c}{\textbf{Anisotropic Rayleigh}} \\
    
    BIC &  $-39.8$  & $-33.9$  & $-30.8$ & $-38.9$ \\
    \hdashline
    \multicolumn{5}{c}{\textbf{Mie (4 species)}} \\
    $m$ &  $1.82(11)$ & $1.91(10)$ & $1.98(13)$ & $1.29(0.09)$ \\
    BIC &  $-49.8$  & $-54.5$  & $-52.1$ & $-48.9$ \\
    \hdashline
    \multicolumn{5}{c}{\textbf{Rayleigh \& Mie}} \\
    $m$ &  $1.24(48)$ & $1.19(09)$ & $1.24(04)$ & $2.22(65)$ \\
    $R/M$ & 31.63(2.75) & 12.16(0.64) & 6.03(1.11) & 2.64(1.07) \\ 
    BIC &  $-61.3$  & $-62.0$  & $-51.2$ & $-49.9$ \\
    \hdashline
    \multicolumn{5}{c}{\textbf{Rayleigh \& Mie \& unpol-MS}} \\
    $m$ &  $1.45(14)$ & $1.15(11)$ & $1.07(07)$ & $1.05(09)$ \\
    $R/M$ & 9.36(17) & 7.04(26) & 3.41(11) & 2.27(08) \\
    MS & $0.05(01)$ & $0.28(10)$ & $0.61(12)$ & $1.64(35)$\\
    BIC &  $-57.9$  & $-59.8$  & $-61.9$ & $-72.0$ \\
    \end{tabular}
    \label{tab:intmodels}
    \tablefoot{Similar to Table~\ref{tab:polmodels} for individual band intensity fits. The BIC values come from intensity data and are not to be compared with those from other tables. The anisotropic Rayleigh $\delta$ parameter has no impact on intensity data.}  
\end{table}

\subsection{Combined scattering processes}

In the past sections, it has become clear that both the main processes of single scattering: Rayleigh and Mie, are present in the DoP and intensity data taken during full Moon at Paranal. Here we combine both processes as in Eq.~\ref{eq:SS}, and we also explore the inclusion of the unpolarized MS effects as in Eq.~\ref{eq:S12_tot}, and fit them to the data. 

\begin{enumerate}[i]
\item \textbf{Polarization fits}:
The resulting fits for individual DoP fits are shown in Table~\ref{tab:polmodels}. We find that the combined models are good representations of the polarization data. This shows that scattering by a set of nonidentical particles results in depolarization -- we use a spherical Rayleigh component, i.e. no depolarization factor-- \citep{bohren98}. However, the simplicity of the anisotropic Rayleigh scattering still prevails, as seen from the lowest BIC values. Nevertheless, the BIC differences in the $RI$ bands between the Rayleigh-Mie-MS model and pure anisotropic Rayleigh scattering are less than 5. It is interesting to note that the ratio of Rayleigh-to-Mie scattering decreases with wavelength in both combined models, an expected trend (see  Figure~1 of e.g. \citealt{patat2011optical}). Additionally, the MS weight increases, again showing that Mie and MS scattering from larger particles are increasingly important at longer wavelengths.

\item \textbf{Intensity fits}:
The intensity fits clearly show that the combined models are preferred according to the BIC (see Table~\ref{tab:intmodels}). The additional unpolarized MS component is significantly preferred for redder bands as seen in the MS parameter and the decreased BIC values. Similarly to the polarization fits, the Mie and MS components become more important at longer wavelengths.

\item \textbf{Polarization and Intensity fits}:
Simultaneous fits to intensity and polarization in each band are shown in Table~\ref{tab:polintmodels} and as solid lines in Figures~\ref{plot_polint1} and~\ref{plot_polint2}. It is evident that these combined processes better explain the data. The joint Rayleigh-Mie-MS scattering is the best of all models by $\Delta$BIC$<30$ for all bands. The resulting Mie contribution increases with wavelength, as does the MS component. 
\end{enumerate}

We note that models including MS are roughly equivalent to models with anisotropic Rayleigh scattering: they both act as a depolarizer. When both effects are taken into account in the fit, one of the two parameters, the MS fraction or the depolarization factor $\delta$, is consistent with zero, showing that they achieve equivalent results.

\begin{table}[h]
    \centering
    \caption{Polarization and intensity fit results}
    \small
    \renewcommand{\arraystretch}{1.2}
    \begin{tabular}{l||c|c|c|c}
                   & \textbf{B}
                   & \textbf{V}
                   & \textbf{R}
                   & \textbf{I} \\
    \hline
     \multicolumn{5}{c}{\textbf{Anisotropic Rayleigh}} \\
    $\delta$ & $0.032(08)$ & $0.035(10)$ & $0.044(14)$ & $0.103(23)$ \\
    BIC &  $-90.3$  & $-79.6$  & $-73.5$ & $-87.1$ \\
    \hdashline
    \multicolumn{5}{c}{\textbf{Mie (4 species)}} \\
    $m$ &  $1.79(07)$ & $1.51(09)$ & $1.45(04)$ & $1.58(02)$ \\
    BIC &  $-96.8$  & $-89.2$  & $-98.2$ & $-111.2$ \\
    \hdashline
    \multicolumn{5}{c}{\textbf{Rayleigh \& Mie}} \\
    $m$ &  $1.75(14)$ & $1.80(08)$ & $1.83(03)$ & $1.87(02)$ \\
    $R/M$ & 11.99(77) & 13.91(39) & 17.49(25) & 14.86(14) \\ 
    BIC &  $-95.5$  & $-104.2$  & $-103.4$ & $-99.3$ \\
    \hdashline
    \multicolumn{5}{c}{\textbf{Rayleigh \& Mie \& unpol-MS}} \\
    $m$ &  $1.05(01)$ & $1.13(01)$ & $1.19(01)$ & $1.18(01)$ \\
    $R/M$ & 5.73(06) & 3.90(04) & 2.62(03) & 1.45(03) \\
    $MS$ & $0.99(02)$ & $1.48(02)$ & $1.73(03)$ & $4.68(06)$\\
    BIC &  $-127.9$  & $-133.3$  & $-137.0$ & $-145.8$ \\
    \end{tabular}
    \label{tab:polintmodels}
    \tablefoot{Similar to Table~\ref{tab:polmodels} for simultaneous intensity and polarization data in each band. BIC values are not to be compared with those of other tables.}
\end{table}

\subsection{Multi-wavelength fits}\label{sec:multiwave}

The models considered in this study have a wavelength dependence as in, e.g. Eq.~\ref{eq:S12_mie}, but refractive indices generally also depend on wavelength. On the other hand, the global ratio of Rayleigh to Mie scattering should be constant if the wavelength dependence is already considered in the models. Similarly, multiple scattering, as given in Eq.~\ref{eq:S12_MS}, already has a measured wavelength dependence, $f(\lambda)$, so that its fraction should be a constant. In this section, we attempt to fit simultaneously all four photometric bands to the various combined models presented earlier. 

\begin{figure}[h]
    \centering
    \includegraphics[width=.495\textwidth]{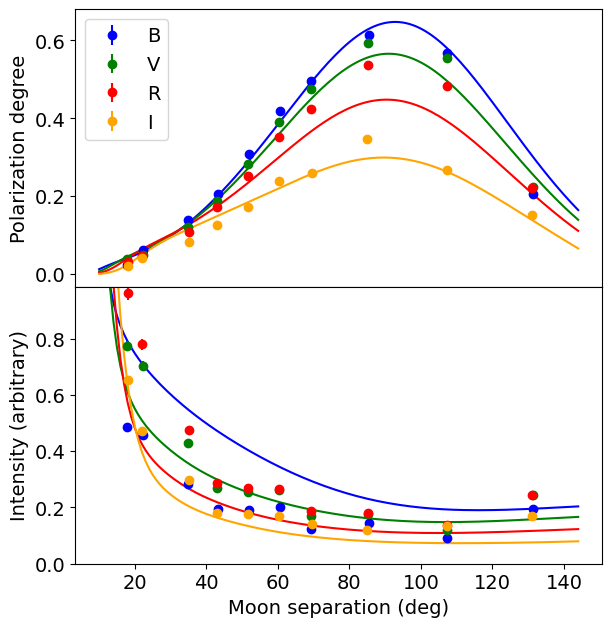}
    \caption{Combined anisotropic Rayleigh $+$ Mie scattering model with 4 aerosol species (with free relative fractions) and a common complex refractive index fitted to simultaneous polarization (\emph{upper}) and intensity (\emph{lower}) multi-wavelength data. Errors are smaller than the points.}
    \label{plot_wavepolint}
\end{figure}

By allowing only a single refractive index for all bands, no model is capable of simultaneously reproducing all bands. If we allow the relative fraction of the four Mie aerosol species to be free in the fit, the results are more promising as shown in Figure~\ref{plot_wavepolint} and Table~\ref{tab:wavepolintmodels}. In particular we find that larger fractions of the tropospheric coarse aerosol is preferred, which is precisely the larger of the four aerosols ($a=0.9 \mu$m) in Table~\ref{tab:aerosols}. Additionally, a non-zero imaginary part of the refractive index also gives more flexibility to the model increasing the wavelength-dependent extinction.

We find that mixed models that include anisotropic Rayleigh scattering with large depolarization factors are preferred over regular Rayleigh with MS contributions. However, the predicted depolarization factors are extremely large ($\delta\sim0.5$) compensating for the lack of MS component. Allowing for both anisotropic Rayleigh \emph{and} MS results in fits where the MS fraction is consistent with zero.  

Despite the overall match of the models as seen in Figure~\ref{plot_wavepolint}, the fits to individual bands of previous sections were generally better. Clearly, more flexibility could be attained by having more than a single refractive index for all four species, by having all six aerosol species contributing, or by allowing the refractive index to change with wavelength. However, having only 10 data points in each band makes the estimation of so many parameters very difficult. In fact, already some of the parameters are poorly constrained, as seen in Table~\ref{tab:wavepolintmodels}.  

\begin{table*}[h]
\begin{threeparttable}
    \centering
    \caption{Multi-wavelength polarization and intensity fit results}
    \small
    \renewcommand{\arraystretch}{1.2}
    \begin{tabular}{l||c|c|c|c|c|c|c}
       Model      & BIC & $R/M$ & MS & $\delta$$^\dag$ & $w_i^{\mathrm{mie}}$(\%)$^*$ & $m_r$ & $m_i$ \\
    \hline
     Ray \& Mie (4 fixed) \& MS& $-262.5$ & 2.21(01) & $0.74(01)$ & -- & -- &  $2.00(01)$ & -- \\
    Ray \& Mie (4 free) \& MS& $-308.4$ & 0.93(17) & $0.90(10)$ & -- & 9(17),29(6),54(19),8 &  $1.58(03)$ & -- \\
    Ray \& Mie (4 fixed, complex) \& MS& $-263.5$ & 1.82(02) & $0.88(02)$ & -- & -- &  $2.00(01)$ & $-0.05(01)$ \\
    Ray \& Mie (4 free, complex) \& MS& $-300.6$ & 0.45(12) & $0.90(21)$ & -- & 8(17),7(4),82(33),3 &  $1.63(54)$ & $-0.04(09)$ \\
    Ani-Ray \& Mie (4 fixed) & $-349.8$ & 1.18(02) & -- & $0.11(01)$ & -- & $1.13(01)$ & -- \\
    Ani-Ray \& Mie (4 free) & $-342.9$ & 1.17(66) & -- & $0.49(01)$ & 16(23),60(12),6(14),18 &  $1.24(22)$ & -- \\
    Ani-Ray \& Mie (4 fixed, complex) & $-379.7$ & 0.82(04) & -- & $0.48(03)$ & -- &  $1.84(25)$ & $-0.93(22)$ \\
    Ani-Ray \& Mie (4 free, complex) & $-402.2$ & 1.36(15) & -- & $0.31(02)$ & 36(14),24(5),37(9),3 &  $1.60(02)$ & $-0.07(01)$ \\
    \end{tabular}
    \label{tab:wavepolintmodels}
    \tablefoot{Similar to Table~\ref{tab:polmodels} for simultaneous intensity and polarization multi-wavelength data. The combined models include Mie components with 4 aerosol species with fixed or free relative fractions ($w_i^{\mathrm{mie}}$) in the fit. A single real or complex refractive index ($m=m_r+im_i$) is assumed for all species. BIC values are not to be compared with those of other tables.}
    \begin{tablenotes}
\small
\item $^{\ast}$ The last relative fraction is not fitted and is constrained from the total of 100\%.
\item $^{\dag}$ The $\delta$ parameter is only present for anisotropic Rayleigh scattering (see section~\ref{sec:aniRay}).
%\item $^{\dagger}$ The models
\end{tablenotes}
\end{threeparttable}
\end{table*}
    
\section{Conclusions}\label{sec:discussion}

We have analyzed full Moon observations from Paranal obtained with FORS2-IPOL and compared them to various scattering models. The degree of polarization of the full moonlit sky peaks near a 90$^{\circ}$ angular separation between the observation target and the Moon, decreasing strongly with wavelength (from 60\% in the $B$ band to 30\% in the $I$ band). Such a pattern is well fit with simple anisotropic Rayleigh scattering models with depolarization factors that increase with wavelength (from 0.03 to 0.11), suggesting an incremental influence of asphericity and/or multiple scattering. The phenomenological framework of \citet{berry2004polarization} also confirms that multiple scattering becomes more dominant at the reddest bands. 

The intensity data are inconsistent with pure Rayleigh scattering and clearly require a Mie scattering component from larger aerosol particles. A combined model of Rayleigh and Mie single scattering plus a multiple scattering contribution, as in \citet{Jones13}, is best suited to the individual band fits to both polarization and intensity data. In these fits, the Mie and MS components become more significant at longer wavelengths.

To fit all wavelengths simultaneously, we require higher contributions of large aerosol species and/or a larger extinction (imaginary) part of the refractive index. It is difficult to constrain these parameters with our current dataset.

Future improvements in the fits would allow the different fractions and refractive indices of the 6 aerosols (Table~\ref{tab:aerosols}) to vary independently. To achieve this, more data are needed, either from new observations or through a careful inspection of archival FORS2 polarization data during moonlit sky. Particularly interesting for the angular distance between neutral points from multiple scattering is data near the Moon (or anti-Moon) positions, which is scarce given its impracticality in scientific observations. 

We stress that if a general physical model explaining all observations is not required and only an estimate or prediction of the polarization pattern is needed, then simple anisotropic Rayleigh scattering provides an excellent description of the polarization. Such a model with depolarization increasing with wavelength (following a power-law dependence) explains full moonlit sky polarization and could be used to correct background polarization of science targets. It remains to be seen whether the model can explain the data of other nights with different Moon phases as well. In particular, at waning/waxing phases, the polarization of the incoming spectrum reflected off the Moon becomes more important and cannot be neglected, as was done in this study. 

Finally, it is worth to mention that linearly polarized laser radars (lidars) have been successfully used to measure the scattering properties of the atmospheric clouds \citep{Scotland71,Sassen91}, and such experiments could be performed at Paranal to understand the characteristics of the aerosols and improve the modeling.

The atmospheric models presented in this study can be used to correct polarimetric data of astronomical sources taken on nights with background moonlight in the sky. In particular, simple anisotropic Rayleigh models with wavelength-dependent depolarization factors reproduce the degree of polarization very well (but not the intensity) and can be used as proxies for the moonlit sky polarization.

\section*{Acknowledgements}
We thank Koraljka Mu\v{z}i\'c and Evgenij S. Zubko for useful discussions. This work is based on observations collected at the European Southern Observatory under ESO programme 106.21L9. S.G.G. acknowledges support from the ESO Scientific Visitor Programme. A.M.G. acknowledges financial support from grant PID2023-152609OA-I00, funded by the Spanish Ministerio de Ciencia, Innovaci\'on y Universidades (MICIU), the Agencia Estatal de Investigaci\'on (AEI, 10.13039/501100011033), and the European Union's European Regional Development Fund (ERDF).

\bibliographystyle{aa} % style aa.bst
\bibliography{bib}

\begin{thebibliography}{51}
\expandafter\ifx\csname natexlab\endcsname\relax\def\natexlab#1{#1}\fi

\bibitem[{{Andersson} {et~al.}(2013){Andersson}, {Piirola}, {De Buizer},
  {Clemens}, {Uomoto}, {Charcos-Llorens}, {Geballe}, {Lazarian}, {Hoang}, \&
  {Vornanen}}]{Andersson13}
{Andersson}, B.~G., {Piirola}, V., {De Buizer}, J., {et~al.} 2013, \apj, 775,
  84

\bibitem[{{Antonucci} \& {Miller}(1985)}]{Antonucci85}
{Antonucci}, R.~R.~J. \& {Miller}, J.~S. 1985, \apj, 297, 621

\bibitem[{{Bagnulo} {et~al.}(2009){Bagnulo}, {Landolfi}, {Landstreet}, {Landi
  Degl'Innocenti}, {Fossati}, \& {Sterzik}}]{Bagnulo09}
{Bagnulo}, S., {Landolfi}, M., {Landstreet}, J.~D., {et~al.} 2009, \pasp, 121,
  993

\bibitem[{Berdyugina {et~al.}(2011)Berdyugina, Berdyugin, Fluri, \&
  Piirola}]{berdyugina2011polarized}
Berdyugina, S., Berdyugin, A., Fluri, D., \& Piirola, V. 2011, The
  Astrophysical Journal Letters, 728, L6

\bibitem[{Berry {et~al.}(2004)Berry, Dennis, \& Lee}]{berry2004polarization}
Berry, M., Dennis, M., \& Lee, R. 2004, New Journal of Physics, 6, 162

\bibitem[{Bohren \& Huffman(1998)}]{bohren98}
Bohren, C.~F. \& Huffman, D.~R. 1998, Absorption and Scattering of Light by
  Small Particles (Wiley)

\bibitem[{{Bulla} {et~al.}(2019){Bulla}, {Covino}, {Kyutoku}, {Tanaka},
  {Maund}, {Patat}, {Toma}, {Wiersema}, {Bruten}, {Jin}, \& {Testa}}]{Bulla19}
{Bulla}, M., {Covino}, S., {Kyutoku}, K., {et~al.} 2019, Nature Astronomy, 3,
  99

\bibitem[{{Chandrasekhar}(1950)}]{Chandrasekhar50}
{Chandrasekhar}, S. 1950, {Radiative transfer.}

\bibitem[{{Chandrasekhar}(1960)}]{Chandrasekhar60}
{Chandrasekhar}, S. 1960, {Radiative transfer}

\bibitem[{{Chandrasekhar} \& {Elbert}(1954)}]{Chandrasekhar54}
{Chandrasekhar}, S. \& {Elbert}, D.~D. 1954, {The illumination and polarization
  of the sunlit sky on Rayleigh scattering}

\bibitem[{{Colina} {et~al.}(1996){Colina}, {Bohlin}, \& {Castelli}}]{Colina96}
{Colina}, L., {Bohlin}, R.~C., \& {Castelli}, F. 1996, \aj, 112, 307

\bibitem[{Coulson(1988)}]{Coulson88}
Coulson, K. 1988, Polarization and Intensity of Light in the Atmosphere,
  Studies in geophysical optics and remote sensing (A. Deepak Pub.)

\bibitem[{David(1847)}]{Brewster1847}
David, B. 1847, Phil Mag, 31, 444

\bibitem[{{Dollfus} \& {Bowell}(1971)}]{Dollfus71}
{Dollfus}, A. \& {Bowell}, E. 1971, \aap, 10, 29

\bibitem[{Elterman(1966)}]{Elterman66}
Elterman, L. 1966, Appl. Opt., 5, 1769

\bibitem[{{Foreman-Mackey} {et~al.}(2013){Foreman-Mackey}, {Hogg}, {Lang}, \&
  {Goodman}}]{emcee13}
{Foreman-Mackey}, D., {Hogg}, D.~W., {Lang}, D., \& {Goodman}, J. 2013, \pasp,
  125, 306

\bibitem[{G{\'a}l {et~al.}(2001)G{\'a}l, Horv{\'a}th, Barta, \&
  Wehner}]{gal2001polarization}
G{\'a}l, J., Horv{\'a}th, G., Barta, A., \& Wehner, R. 2001, Journal of
  Geophysical Research: Atmospheres, 106, 22647

\bibitem[{Gandorfer(2000)}]{Gandorfer00}
Gandorfer, A. 2000, Social Studies of Science

\bibitem[{Gandorfer(2002)}]{Gandorfer02}
Gandorfer, A. 2002, The Second Solar Spectrum, Vol. II: 3910{\AA} to 4630{\AA}

\bibitem[{{Gonz{\'a}lez-Gait{\'a}n} {et~al.}(2020){Gonz{\'a}lez-Gait{\'a}n},
  {Mour{\~a}o}, {Patat}, {Anderson}, {Cikota}, {Wiersema}, {Higgins}, \&
  {Silva}}]{Gonzalez-Gaitan20}
{Gonz{\'a}lez-Gait{\'a}n}, S., {Mour{\~a}o}, A.~M., {Patat}, F., {et~al.} 2020,
  \aap, 634, A70

\bibitem[{{Hansen} \& {Travis}(1974)}]{Hansen1974}
{Hansen}, J.~E. \& {Travis}, L.~D. 1974, \ssr, 16, 527

\bibitem[{{Harrington} {et~al.}(2017){Harrington}, {Kuhn}, \&
  {Ariste}}]{Harrington17}
{Harrington}, D.~M., {Kuhn}, J.~R., \& {Ariste}, A.~L. 2017, Journal of
  Astronomical Telescopes, Instruments, and Systems, 3, 018001

\bibitem[{Horv{\'a}th {et~al.}(2002)Horv{\'a}th, Bern{\'a}th, Suhai, Barta, \&
  Wehner}]{horvath2002first}
Horv{\'a}th, G., Bern{\'a}th, B., Suhai, B., Barta, A., \& Wehner, R. 2002,
  JOSA A, 19, 2085

\bibitem[{{Jones} {et~al.}(2013){Jones}, {Noll}, {Kausch}, {Szyszka}, \&
  {Kimeswenger}}]{Jones13}
{Jones}, A., {Noll}, S., {Kausch}, W., {Szyszka}, C., \& {Kimeswenger}, S.
  2013, \aap, 560, A91

\bibitem[{{Jones} {et~al.}(2019){Jones}, {Noll}, {Kausch}, {Unterguggenberger},
  {Szyszka}, \& {Kimeswenger}}]{Jones19}
{Jones}, A., {Noll}, S., {Kausch}, W., {et~al.} 2019, \aap, 624, A39

\bibitem[{{Kohan}(1962)}]{Kohan62}
{Kohan}, E.~K. 1962, in The Moon, ed. Z.~{Kopal} \& Z.~K. {Mikhailov}, Vol.~14,
  453--462

\bibitem[{{Leonard} {et~al.}(2001){Leonard}, {Filippenko}, {Ardila}, \&
  {Brotherton}}]{Leonard01}
{Leonard}, D.~C., {Filippenko}, A.~V., {Ardila}, D.~R., \& {Brotherton}, M.~S.
  2001, \apj, 553, 861

\bibitem[{Liou(2002)}]{liou2002introduction}
Liou, K.-N. 2002, An introduction to atmospheric radiation, Vol.~84 (Elsevier)

\bibitem[{Liu {et~al.}(2020)Liu, Fan, \& Wang}]{liu2020solar}
Liu, B., Fan, Z., \& Wang, X. 2020, IEEE Access, 8, 56720

\bibitem[{Madhusudhan \& Burrows(2012)}]{madhusudhan2012analytic}
Madhusudhan, N. \& Burrows, A. 2012, The Astrophysical Journal, 747, 25

\bibitem[{Narasimhan \& Nayar(2003)}]{narasimhan2003shedding}
Narasimhan, S.~G. \& Nayar, S.~K. 2003, in 2003 IEEE Computer Society
  Conference on Computer Vision and Pattern Recognition, 2003. Proceedings.,
  Vol.~1, IEEE, I--I

\bibitem[{Newville {et~al.}(2016)Newville, Stensitzki, Allen, Rawlik,
  Ingargiola, \& Nelson}]{newville2016lmfit}
Newville, M., Stensitzki, T., Allen, D.~B., {et~al.} 2016, Astrophysics Source
  Code Library, ascl

\bibitem[{Patat {et~al.}(2011)Patat, Moehler, O’Brien, Pompei, Bensby,
  Carraro, de~Ugarte~Postigo, Fox, Gavignaud, James,
  {et~al.}}]{patat2011optical}
Patat, F., Moehler, S., O’Brien, K., {et~al.} 2011, Astronomy \&
  Astrophysics, 527, A91

\bibitem[{{Penndorf}(1957)}]{Penndorf57}
{Penndorf}, R. 1957, Journal of the Optical Society of America (1917-1983), 47,
  176

\bibitem[{Prahl(2023)}]{miepython}
Prahl, S. 2023, miepython

\bibitem[{Rayleigh(1871)}]{Rayleigh1871}
Rayleigh, L. 1871, Phil Mag, 41, 274

\bibitem[{{Rino-Silvestre} {et~al.}(2025){Rino-Silvestre},
  {Gonz{\'a}lez-Gait{\'a}n}, {Mour{\~a}o}, {Duarte}, \&
  {Pereira}}]{Rino-Silvestre25}
{Rino-Silvestre}, J., {Gonz{\'a}lez-Gait{\'a}n}, S., {Mour{\~a}o}, A.,
  {Duarte}, J., \& {Pereira}, B. 2025, arXiv e-prints, arXiv:2502.09875

\bibitem[{Sassen(1991)}]{Sassen91}
Sassen, K. 1991, Bulletin of the American Meteorological Society, 72, 1848

\bibitem[{{Scarrott} {et~al.}(1991){Scarrott}, {Rolph}, {Wolstencroft}, \&
  {Tadhunter}}]{Scarrott91}
{Scarrott}, S.~M., {Rolph}, C.~D., {Wolstencroft}, R.~W., \& {Tadhunter}, C.~N.
  1991, \mnras, 249, 16P

\bibitem[{Schotland {et~al.}(1971)Schotland, Sassen, \& Stone}]{Scotland71}
Schotland, R.~M., Sassen, K., \& Stone, R. 1971, Journal of Applied Meteorology
  and Climatology, 10, 1011

\bibitem[{{Steinacker} {et~al.}(2013){Steinacker}, {Baes}, \&
  {Gordon}}]{Steinacker13}
{Steinacker}, J., {Baes}, M., \& {Gordon}, K.~D. 2013, \araa, 51, 63

\bibitem[{{Stenflo}(2005)}]{Stenflo05}
{Stenflo}, J.~O. 2005, \aap, 429, 713

\bibitem[{{Stenflo} \& {Keller}(1997)}]{Stenflo97}
{Stenflo}, J.~O. \& {Keller}, C.~U. 1997, \aap, 321, 927

\bibitem[{{Stenflo} {et~al.}(2000){Stenflo}, {Keller}, \&
  {Gandorfer}}]{Stenflo00}
{Stenflo}, J.~O., {Keller}, C.~U., \& {Gandorfer}, A. 2000, \aap, 355, 789

\bibitem[{{Stokes}(1851)}]{Stokes1851}
{Stokes}, G.~G. 1851, Transactions of the Cambridge Philosophical Society, 9,
  399

\bibitem[{Stuke(2016)}]{phdthesis}
Stuke, S. 2016, PhD thesis

\bibitem[{Wang {et~al.}(2016)Wang, Gao, Fan, \& Roberts}]{wang2016analytical}
Wang, X., Gao, J., Fan, Z., \& Roberts, N.~W. 2016, Journal of Optics, 18,
  065601

\bibitem[{Williams \& Warneck(2012)}]{Williams12}
Williams, J. \& Warneck, P. 2012, The Atmospheric Chemist´s Companion

\bibitem[{{Wolf}(2003)}]{Wolf03}
{Wolf}, S. 2003, \apj, 582, 859

\bibitem[{{Wolff}(1975)}]{Wolff75}
{Wolff}, M. 1975, \ao, 14, 1395

\bibitem[{{Yue} {et~al.}(1994){Yue}, {Poole}, {Wang}, \& {Chiou}}]{Yue94}
{Yue}, G.~K., {Poole}, L.~R., {Wang}, P.~H., \& {Chiou}, E.~W. 1994, \jgr, 99,
  3727

\end{thebibliography}

\appendix

\section{\label{ap:polang} Polarization Angle}

\begin{figure}[h]
    \centering
    \includegraphics[width=.49\textwidth]{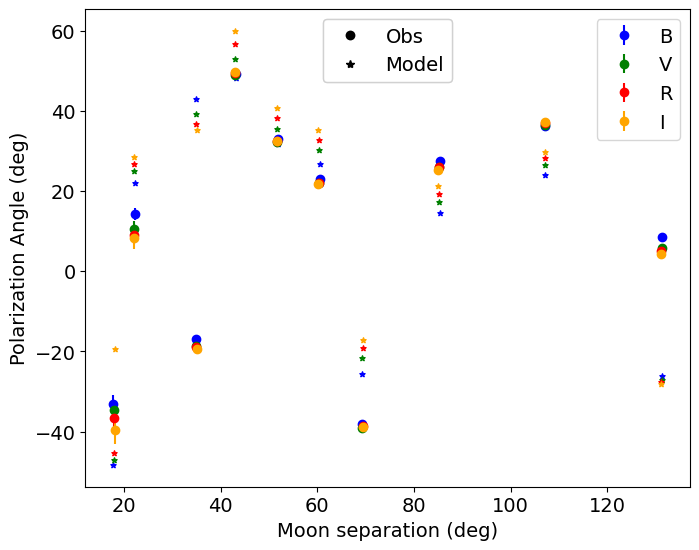}
    \caption{Observed polarization angle (points) versus moon separation compared to the expected angle (stars) obtained from simple rotations of the scattering plane (Eq.~\ref{eq:muller-rotation} and Eq.~\ref{eq:polang}).}
    \label{fig:polangle}
\end{figure}

We present in Fig.~\ref{fig:polangle} the measured polarization angle at different angular separations from the Moon in different filters. The stars represent the expected angle obtained from incoming unpolarized light that is scattered in the atmosphere. In such a case, the polarization angle is only dependent on the scattering plane with respect to the incoming light and the observer (see Eq.~\ref{eq:muller-rotation} and Eq.~\ref{eq:polang}), and does not depend on the atmospheric particles nor on the wavelength of the light. The variations seen in the model among different filters arise because of the different times of observation that influence the relative Moon-scattering-observer angles. We roughly confirm here the expected angles, although there are clear discrepancies for certain observations, notably at $\gamma\sim35^{\circ}$ and $\gamma\sim130^{\circ}$.

\section{\label{sec:atm} Atmospheric considerations}

Although no specific polarimetric models exist for Paranal, general analytic models include the depolarization effects of the turbidity and optical thickness of the atmosphere based on empirical corrections. In the study of \citet{wang2016analytical}, the stronger depolarization effects near the horizon compared with the zenith are modeled by introducing a horizon correction factor, $E$, that has a minimum value at the zenith ($\theta=0^{\circ}$) and decreases towards the horizon ($\theta=90^{\circ}$). So the correction becomes: $E(\theta) = cos(\theta)^{\frac{1}{N}}$, 
where $N$ is a control parameter used as a free parameter to fit. In addition, they include a correction factor related to the turbulence of the atmosphere. We consider here the "seeing" ($S$) parameter as an indicator of the turbulent air flows in the atmosphere. The turbulence generates a gradual depolarization, $M(S)$, that follows an exponential fall off: $M(S) = e^{-\frac{S}{k_1}+k_2}$,
where the constants $k_1$ and $k_2$ are free parameters used to fit the relationship. 

Considering these two effects, the corrected polarization degree is described by:

\begin{equation}
    p_{\mathrm{corr}} = M(S) E(\theta) \times p = e^{-\frac{S}{k_1}+k_2} \cos(\theta)^{\frac{1}{N}} \times p \;.
    \label{eq:mix}
\end{equation}

\begin{figure}[h]
    \centering
    \includegraphics[width=.5\textwidth]{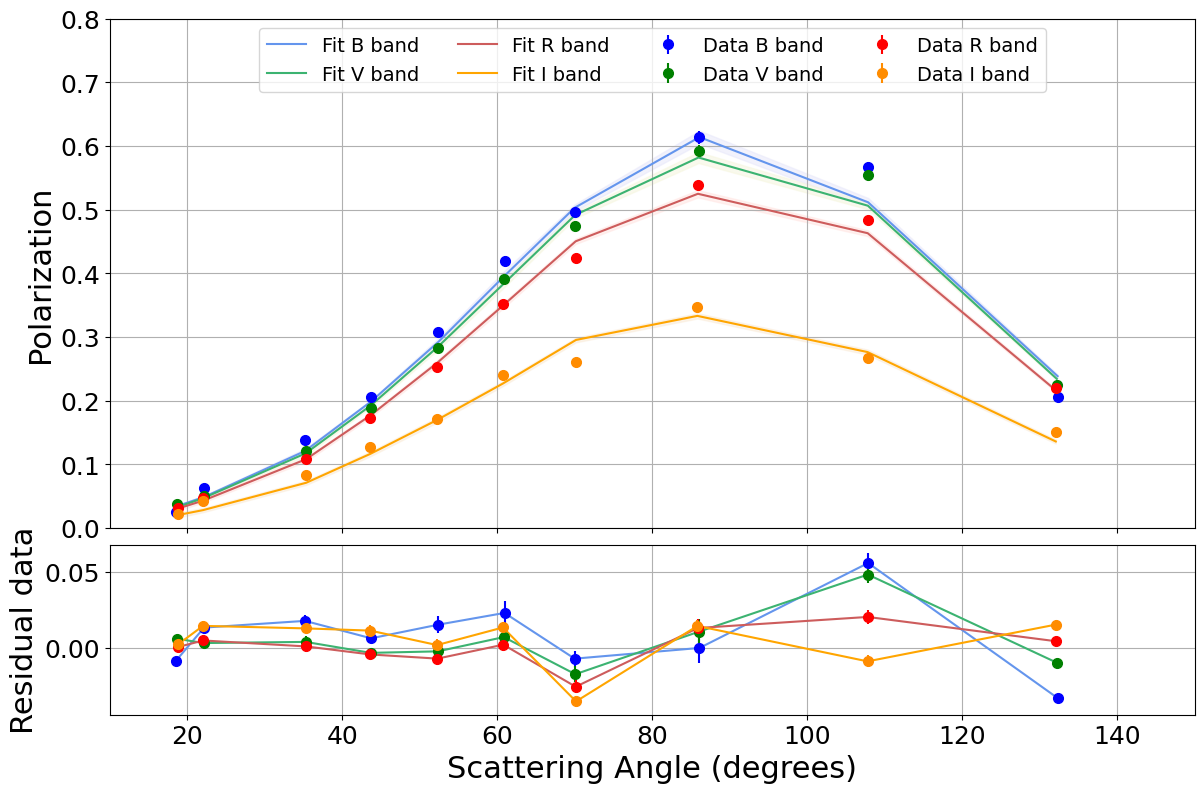}
    \caption{Fits considering the atmospheric corrections of equation \ref{eq:mix} for the single Rayleigh scattering model.}
    \label{plot22}
\end{figure}

Using the formalism of Eq.~\ref{eq:mix} to correct a pure Rayleigh scattering model, we obtain the fits shown in Figure~\ref{plot22} which reveal improvements in the $\chi^2$ with respect to pure Rayleigh, yet the BIC values punish the increase of free parameters. It is encouraging to see that such empirical corrections may indeed improve the fits. This might be of particular importance with larger, more heterogeneous datasets.

\section{\label{sec:Apthree}Twilight data}

We analyze the Sun polarization contribution from twilight data -- which corresponds to the Sun being 18$^{\circ}$ below the horizon or higher. To consider both sources, the Moon and the Sun, their Stokes vectors are added in the Muller calculations, and the albedo is considered to be the sum of each light source contribution.

\begin{figure}[h]
    \centering
    \includegraphics[width=.5\textwidth]{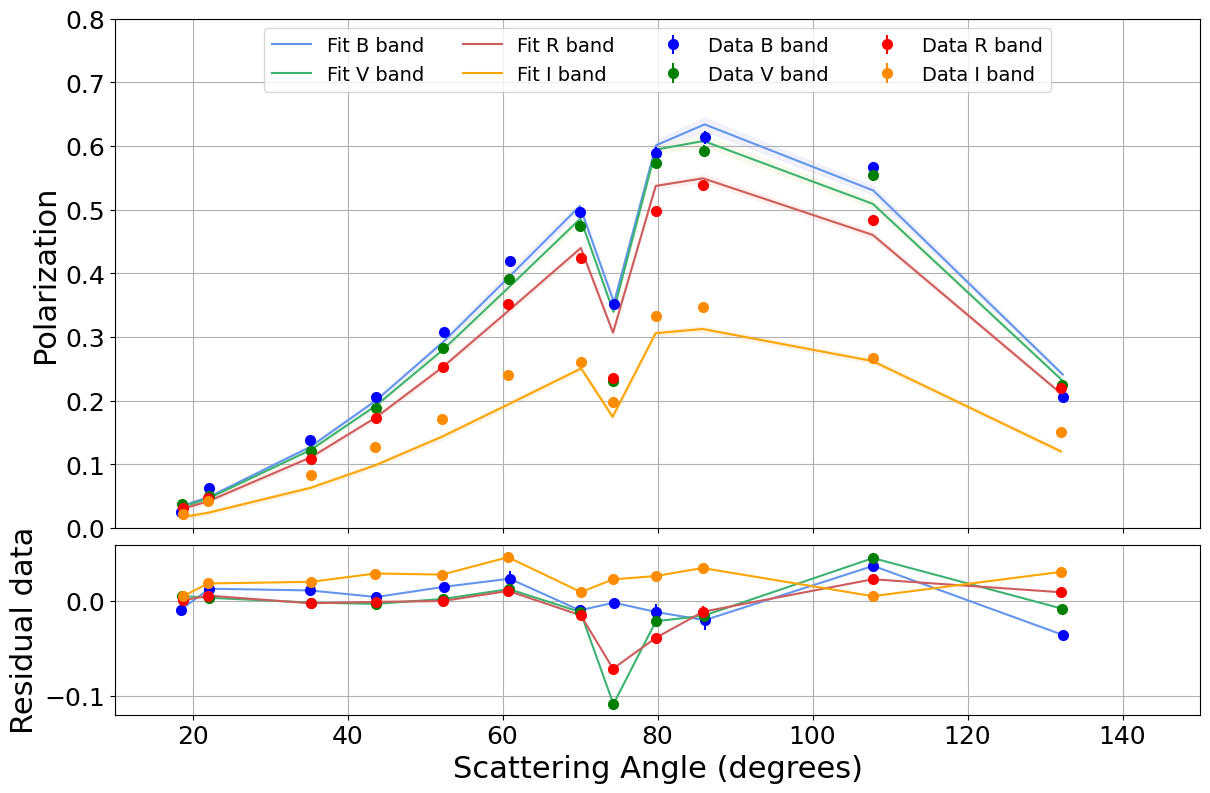}
    \caption{Fit of the Rayleigh model considering the polarization with the contribution from the Sun.}
    \label{plot14}
\end{figure}

We assume pure Rayleigh scattering and a relative light fraction parameter $f$ between both sources. The Sun-Moon model shown in Figure~\ref{plot14} fits the data well, including the dip due to twilight. The fit values for the fraction parameter $f$ are shown in Table~\ref{tab:fparameter}.  

\begin{table}
    \centering
    \caption{Fit results for parameter $f$ in each band}
    \small
    \renewcommand{\arraystretch}{1.2}
    \begin{tabular}{l||c|c|c|c}
                   & \textbf{B}
                   & \textbf{V}
                   & \textbf{R}
                   & \textbf{I} \\
    \hline
    $f(10^{-10})$ & $5.4\pm0.8$ & $0.24\pm0.004 $ & $0.11\pm0.02$ & $18\pm12$ \\
    \end{tabular}
    \label{tab:fparameter}
    \tablefoot{Best-fit values of the parameter $f$ and associated uncertainties for each band.}
\end{table}

\section{MCMC sampling}\label{ap:mcmc}
\begin{figure}[h]
    \centering
    \includegraphics[width=.49\textwidth]{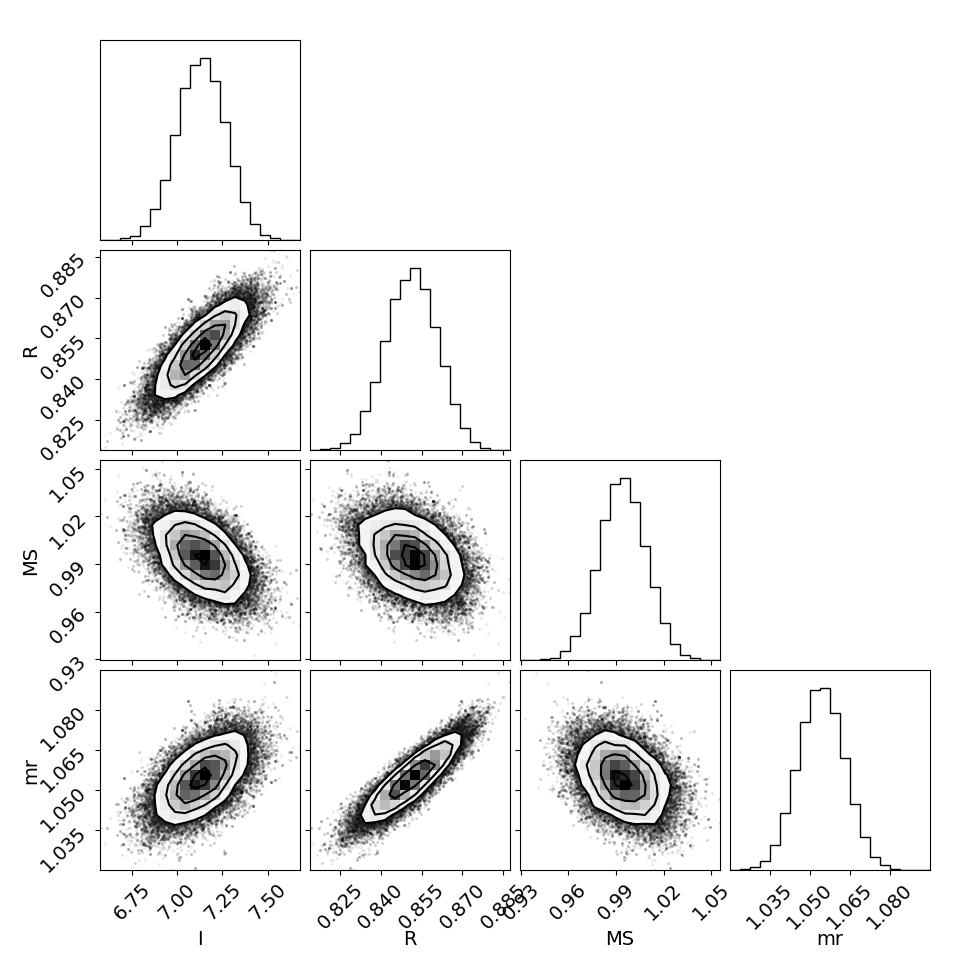}
    \caption{Corner plot of the four free parameters of an isotropic Rayleigh-Mie-MS model (4 Mie species with fixed fractions and unpolarized MS) fitted to $B$-band: the real refractive index $m_r$, the intensity normalisation parameter $I$, the Rayleigh component $R$ (the Mie $M$ component is given by $M=1-R$) and the Multiple Scattering factor $MS$.}
    \label{fig:mcmc}
\end{figure}

\begin{figure}[h]
    \centering
    \includegraphics[width=.49\textwidth]{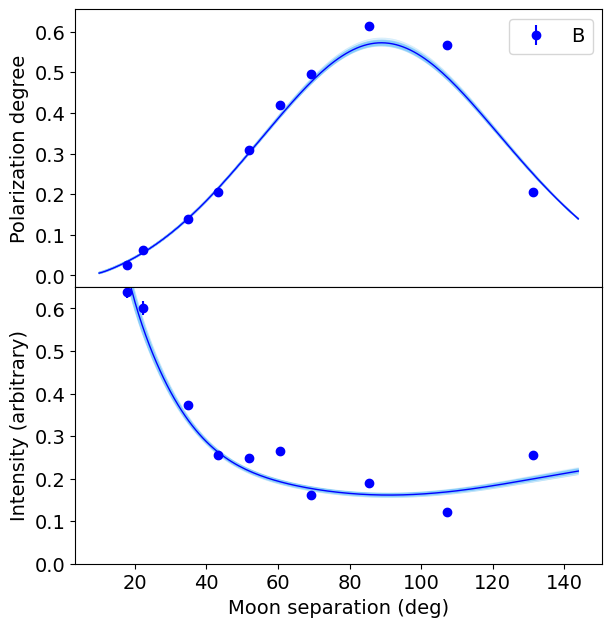}
    \caption{Rayleigh-Mie-MS fit (blue lines) to $B$-band polarization (upper) and intensity (lower) data (uncertainties are smaller than data points). The dark and light blue lines represent the median and 200 realizations drawn from the posterior distribution of the MCMC, respectively.}
    \label{fig:fitmcmc}
\end{figure}

We show here an example of the MCMC sampling results. In each case, we sample the parameter space with 132 walkers and 800 iterations with a burn-in of 200 iterations. For a few cases, we extended the sampling to 1200 and burn-in to 600 iterations when the chains had not converged. The priors used are flat for all parameters. In Fig.~\ref{fig:mcmc}, we show the corner plot for a Rayleigh-Mie-MS fit to $B$-band and its respective fit in Fig.~\ref{fig:fitmcmc} with 200 realizations from the posterior distribution shown in light blue. We note that in all cases the median values of the posterior distribution are consistent within 1$\sigma$ with the Levenberg-Marquardt minimization results. 

\section{Polarization wavelength-dependence}\label{ap:powlaw}

\begin{figure}[h]
    \centering
    \includegraphics[width=.49\textwidth]{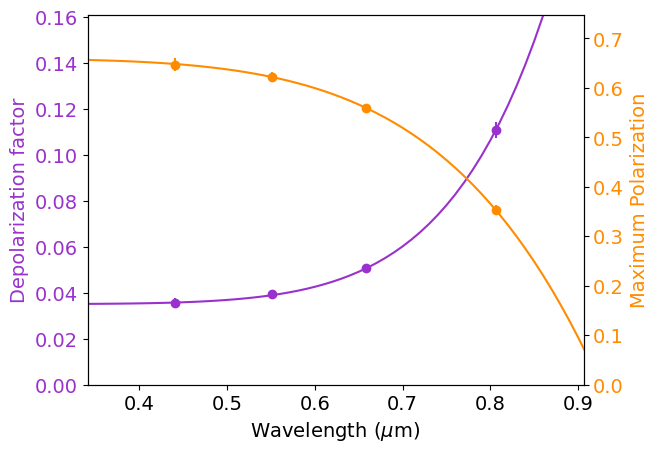}
    \caption{Wavelength-dependence of the depolarization factor, $\delta$, for anisotropic Rayleigh scattering (purple) and a normalization factor, $A$, multiplying a pure Rayleigh model (uncertainties are smaller than data points). The lines are power-law fits to the data (see Table~\ref{tab:powlaw}.}
    \label{fig:powlaw}
\end{figure}

In this section we investigate the wavelength dependence of the polarization data. Since polarization only is very well modeled with only Rayleigh scattering, we study the obtained depolarization factor $\delta(\lambda)$ from the anisotropic Rayleigh fits (Eq.~\ref{eq:ray-depol}). Additionally, we also consider a simple constant, $A(\lambda)$, that multiplies the Rayleigh scattering in each band (Eq.~\ref{eq:rayleigh}). We fit both relations to a power-law of the form: $p=B\lambda^n+C$, as shown in Fig.~\ref{fig:powlaw} and Table~\ref{tab:powlaw}. The recovered power laws are in disagreement with previous claims of $n=-4$ \citep{Andersson13}. This stems probably from the different combination at various wavelengths of several atmospheric species and scattering processes (single Rayleigh and Mie scattering, and multiple scattering).

\begin{table}
    \centering
    \caption{Parameters of power-law fit ($p=B\lambda^n+C$) to $\delta(\lambda)$ and $A(\lambda)$.}
    \small
    \renewcommand{\arraystretch}{1.2}
    \begin{tabular}{l||c|c|c}
                   & \textbf{B}
                   & \textbf{n}
                   & \textbf{C}\\
    \hline
    $\delta(\lambda)$ & $0.41(02)$ & $7.82(21)$ & $0.0351(01)$ \\
    $A(\lambda)$ & $-1.01(01)$ & $5.55(05)$ & $0.6591(14)$ \\
    \end{tabular}
    \label{tab:powlaw}
    %\tablefoot{Best-fit values of the parameter $f$ and associated uncertainties for each band.}
\end{table}

\section{\label{ap:data}Observational Data}
We present in Tables~\ref{Bobs-table},~\ref{Vobs-table},~\ref{Robs-table} and~\ref{Iobs-table} the $BVRI$ data obtained with FOR2-IPOL. In the table columns, \emph{Field} refers to the name of the observed region, \emph{RA} and \emph{DEC} indicate the right ascension and declination of the field center in degrees, $\overline{t}_{\text{obs}}$ corresponds to the average universal time during the observation, \emph{Exp} is the exposure time in seconds, \emph{Seeing} denotes the atmospheric seeing conditions during the observation in arcseconds, \emph{A} represents the albedo at the time of observation expressed as a percentage, $\gamma$ is the scattering angle in degrees, \emph{AOP} refers to the angle of polarization in degrees, and \emph{DOP} is the degree of polarization in percentages. All data are presented regardless of the observation conditions. We note that the field CT1 was observed with eigh HWP angle positions \citep[see][]{Gonzalez-Gaitan20}.

\begin{table*}
    \centering
    \caption{Summary of the observational data obtained with the B filter.}
    \begin{tabular}{|c|c|c|c|c|c|c|c|c|c|}
    \hline
        Field & RA & DEC & $\overline{t}_{obs}$ (UTC) & Exp & Seeing  & A ($\%$) & $\gamma$ ($^{\circ}$) & AOP ($^{\circ}$) & DOP ($\%$)\\ \hline
        F60 & 89.47221 & -13.62131 & 2021-01-26T00:42:47 & 186 & 1.07  & 8.7 & 39.16 & 7.25$\pm$0.38 & 12.1$\pm$ 0.3 \\ \hline
        F9 & 66.66121 & 24.65011 & 2021-01-26T01:34:23 & 292 & 1.67 & 8.8 & 24.54 & -4.29$\pm$1.62 & 6.9$\pm$ 0.2 \\ \hline
        F58 & 75.41671 & -48.90769 & 2021-01-27T00:58:01 & 129 & 0.95 & 11.1 & 79.81 & 32.21$\pm$0.62 & 59.0$\pm$ 0.8 \\ \hline
        F55 & 54.02579 & -20.96981 & 2021-01-27T01:18:03 & 266 & 0.79 & 11.0 & 69.97 & -36.65$\pm$0.60 & 49.6$\pm$ 0.6 \\ \hline
        F10 & 71.42771 & 17.19769 & 2021-01-27T02:10:40 & 179 & 1.55 & 11.1 & 35.17 & -15.31$\pm$0.96 & 13.9$\pm$0.4 \\ \hline
        F11 & 82.95929 & 12.62319 & 2021-01-26T02:38:12 & 140 & 1.33 & 8.9 & 16.58 & -17.38$\pm$3.02 & 2.5$\pm$0.2 \\ \hline
        F17 & 133.15200 & 24.72839 & 2021-01-27T03:19:41 & 162 & 0.87 & 11.2 & 22.15 & 15.84$\pm$1.49 & 6.2$\pm$0.2 \\ \hline
        9090-7 & 138.00100 & -7.84645 & 2021-01-27T04:01:07 & 196 & 0.94 & 11.3 & 43.71 & -39.26$\pm$0.33 & 20.5$\pm$0.3 \\ \hline
        F64 & 156.58800 & -0.12736 & 2021-01-27T04:46:18 & 218 & 0.79 & 11.3 & 52.39 & 34.61$\pm$0.45 & 30.8$\pm$0.6 \\ \hline
        F23 & 173.41300 & 13.44331 & 2021-01-27T05:37:26 & 242 & 0.70 & 11.4 & 60.92 & 24.65$\pm$0.52 & 41.9$\pm$0.8 \\ \hline
        CT2 & 194.38792 & -2.38783 & 2021-01-27T06:31:12 & 298 & 1.13 & 11.5 & 86.08 & 28.95$\pm$0.48 & 61.4$\pm$1.0 \\ \hline
        F74 & 235.73100 & -34.11100 & 2021-01-27T07:31:53 & 209 & 0.63 & 11.5 & 132.28 & 10.07$\pm$0.84 & 20.6$\pm$0.3 \\ \hline
        F65 & 166.39100 & -77.77269 & 2021-01-27T08:19:28 & 123 & 0.69 & 11.6 & 107.84 & 37.92$\pm$0.48 & 56.7$\pm$0.6 \\ \hline
        1101-264 & 166.20418 & -27.00247 & 2021-01-27T08:56:44 & 124 & 0.71 & 11.7 & 74.38 & 42.70$\pm$0.53 & 35.3$\pm$0.3 \\ \hline
        CT1 & 151.11239 & -2.31674 & 2017-03-12T01:05:50 & 200 & 0.89 & 12.8 & 18.51 & -33.22$\pm$2.22 & 2.5$\pm$0.2 \\ \hline
    \end{tabular}\label{Bobs-table}
\end{table*}

\begin{table*}
    \centering
    \caption{Summary of the observational data obtained with the V filter.}
    \begin{tabular}{|c|c|c|c|c|c|c|c|c|c|}
    \hline
        Field & RA & DEC & $\overline{t}_{obs}$ (UTC) & Exp & Seeing & A ($\%$) & $\gamma$ ($^{\circ}$) & AOP ($^{\circ}$) & DOP ($\%$) \\ \hline
        F60 & 89.47221 & -13.62131 & 2021-01-26T00:56:00 & 86 & 0.96 & 10.4 & 39.16 & 5.85$\pm$0.37 & 12.3$\pm$0.3 \\ \hline
        F9 & 66.66121 & 24.65011 & 2021-01-26T01:53:13 & 136 & 1.35 & 10.5 & 24.64 & -6.89$\pm$1.93 & 5.8$\pm$0.2 \\ \hline
        F58 & 75.41671 & -48.90769 & 2021-01-27T00:48:59 & 60 & 0.90 & 13.0 & 79.79 & 31.78$\pm$0.50 & 57.3$\pm$0.6 \\ \hline
        F55 & 54.02579 & -20.96981 & 2021-01-27T01:35:35 & 124 & 0.74 & 13.1 & 70.05 & -37.19$\pm$0.50 & 47.4$\pm$0.5 \\ \hline
        F10 & 71.42771 & 17.19769 & 2021-01-27T02:23:27 & 83 & 1.62 & 13.2 & 35.24 & -16.90$\pm$1.02 & 12.1$\pm$0.4 \\ \hline
        F11 & 82.95929 & 12.62319 & 2021-01-26T02:57:59 & 70 & 2.03 & 10.6 & 16.65 & -23.89$\pm$4.20 & 1.84$\pm$0.3 \\ \hline
        F17 & 133.15200 & 24.72839 & 2021-01-27T03:31:34 & 75 & 0.81 & 13.3 & 22.09 & 12.52$\pm$2.00 & 5.0$\pm$0.2 \\ \hline
        9090-7 & 138.00100 & -7.84645 & 2021-01-27T04:14:54 & 91 & 0.91 & 13.3 & 43.65 & -39.14$\pm$0.20 & 18.9$\pm$0.3 \\ \hline
        F64 & 156.58800 & -0.12736 & 2021-01-27T05:01:14 & 101 & 0.69 & 13.4 & 52.31 & 34.18$\pm$0.31 & 28.2$\pm$0.5 \\ \hline
        F23 & 173.41300 & 13.44331 & 2021-01-27T05:53:36 & 112 & 0.68 & 13.5 & 60.82 & 23.86$\pm$0.44 & 39.1$\pm$0.6 \\ \hline
        CT2 & 194.38792 & -2.38783 & 2021-01-27T06:50:27 & 139 & 1.51 & 13.6 & 85.95 & 27.89$\pm$0.37 & 59.2$\pm$0.9 \\ \hline
        F74 & 235.73100 & -34.11100 & 2021-01-27T07:46:16 & 97 & 0.65 & 13.7& 132.18 & 7.55$\pm$0.80 & 22.4$\pm$0.2 \\ \hline
        F65 & 166.39100 & -77.77269 & 2021-01-27T08:29:18 & 57 & 0.69 & 13.7 & 107.80 & 38.20$\pm$0.33 & 55.4$\pm$0.5 \\ \hline
        1101-264 & 166.20418 & -27.00247 & 2021-01-27T09:06:40 & 57 & 0.77 & 13.8 & 74.31 & -44.78$\pm$0.26 & 23.1$\pm$0.3 \\ \hline
        CT2 & 151.11239 & -2.31674 & 2017-03-12T01:41:48 & 120 & 1.22 & 15.1 & 18.65 & -34.68$\pm$1.80 & 3.8$\pm$0.3 \\ \hline
    \end{tabular}\label{Vobs-table}
\end{table*}

\begin{table*}
    \centering
    \caption{Summary of the observational data obtained with the R filter.}
    \begin{tabular}{|c|c|c|c|c|c|c|c|c|c|}
    \hline
        Field & RA & DEC & $\overline{t}_{obs}$ (UTC) & Exp & Seeing & A ($\%$) & $\gamma$ ($^{\circ}$) & AOP ($^{\circ}$) & DOP ($\%$) \\ \hline
        F60 & 89.47221 & -13.62131 & 2021-01-26T01:04:19 & 60 & 1.08 & 11.7 & 39.17 & 3.60$\pm$0.60 & 11.3$\pm$0.3 \\ \hline
        F9 & 66.66121 & 24.65011 & 2021-01-26T02:04:24 & 94 & 1.06 & 11.9 & 24.69 & -8.96$\pm$1.88 & 5.9$\pm$0.3 \\ \hline
        F58 & 75.41671 & -48.90769 & 2021-01-27T00:40:38 & 41 & 0.89 & 14.6 & 79.76 & 30.98$\pm$0.37 & 49.8$\pm$0.4 \\ \hline
        F55 & 54.02579 & -20.96981 & 2021-01-27T01:46:07 & 86 & 0.79 & 14.7 & 70.10 & -39.88$\pm$0.38 & 42.5$\pm$0.4 \\ \hline
        F10 & 71.42771 & 17.19769 & 2021-01-27T02:32:09 & 58 & 1.60 & 14.8 & 35.29 & -20.01$\pm$1.00 & 10.9$\pm$0.4 \\ \hline
        F11 & 82.95929 & 12.62319 & 2021-01-27T02:53:07 & 30 & 0.81 & 14.8 & 27.29 & -31.12$\pm$1.20 & 7.4$\pm$0.4 \\ \hline
        F17 & 133.15200 & 24.72839 & 2021-01-27T03:39:10 & 52 & 0.82 & 14.9 & 22.05 & 8.12$\pm$2.34 & 4.8$\pm$0.3 \\ \hline
        9090-7 & 138.00100 & -7.84645 & 2021-01-27T04:23:28 & 63 & 0.86 & 15.0 & 43.62 & -41.60$\pm$0.28 & 17.2$\pm$0.3 \\ \hline
        F64 & 156.58800 & -0.12736 & 2021-01-27T05:10:25 & 70 & 0.73 & 15.1 & 52.26 & 31.42$\pm$0.30 & 25.3$\pm$0.4 \\ \hline
        F23 & 173.41300 & 13.44331 & 2021-01-27T06:03:37 & 88 & 0.74 & 15.2 & 60.76 & 20.95$\pm$0.38 & 35.2$\pm$0.4 \\ \hline
        CT2 & 194.38792 & -2.38783 & 2021-01-27T07:02:07 & 106 & 1.75 & 15.3 & 85.87 & 24.78$\pm$0.28 & 53.8$\pm$0.6 \\ \hline
        F74 & 235.73100 & -34.11100 & 2021-01-27T07:55:24 & 77 & 0.62 & 15.4 & 132.11 & 3.92$\pm$0.65 & 22.0$\pm$0.3 \\ \hline
        F65 & 166.39100 & -77.77269 & 2021-01-27T08:35:52 & 39 & 0.69 & 15.4 & 107.78 & 35.85$\pm$0.18 & 48.3$\pm$0.5 \\ \hline
        1101-264 & 166.20418 & -27.00247 & 2021-01-27T09:13:25 & 45 & 0.82 & 15.5 & 74.25 & -44.78$\pm$0.21 & 23.6$\pm$0.3 \\ \hline
        CT3 & 151.11239 & -2.31674 & 2017-03-12T02:03:41 & 100 & 1.06 & 17.0 & 18.73 & -36.70$\pm$2.40 & 3.1$\pm$0.3\\ \hline
    \end{tabular}\label{Robs-table}
\end{table*}

\begin{table*}
    \centering
    \caption{Summary of the observational data obtained with the I filter.}
    \begin{tabular}{|c|c|c|c|c|c|c|c|c|c|}
    \hline
        Field & RA & DEC & $\overline{t}_{obs}$ (UTC) & Exp & Seeing & A ($\%$) & $\gamma$ ($^{\circ}$) & AOP ($^{\circ}$) & DOP ($\%$) \\ \hline
        F60 & 89.47221 & -13.62131 & 2021-01-26T01:11:17 & 53 & 1.05 & 13.2 & 39.19 & 1.90$\pm$1.20 & 8.8$\pm$0.3 \\ \hline
        F9 & 66.66121 & 24.65011 & 2021-01-26T02:13:27 & 83 & 1.97 & 13.3 & 24.74 & -10.54$\pm$2.18 & 4.9$\pm$0.3 \\ \hline
        F58 & 75.41671 & -48.90769 & 2021-01-27T00:36:55 & 37 & 0.95 & 16.3 & 79.75 & 38.31$\pm$0.277 & 33.3$\pm$0.4 \\ \hline
        F55 &  54.02579 & -20.96981 & 2021-01-27T01:54:44 & 76 & 0.79 & 16.5 & 70.14 & -41.67$\pm$0.31 & 26.0$\pm$0.4 \\ \hline
        F10 & 71.42771 & 17.19769 & 2021-01-27T02:38:26 & 51 & 1.61 & 16.6 & 35.32 & -22.21$\pm$1.18 & 8.4$\pm$0.3 \\ \hline
        F11 & 82.95929 & 12.62319 & 2021-01-26T03:01:48 & 25 & 0.87 & 13.4 & 16.67 & -34.07$\pm$1.31 & 7.0$\pm$0.4\\ \hline
        F17 & 133.15200 & 24.72839 & 2021-01-27T03:45:32 & 46 & 0.91 & 16.7 & 22.02 & 5.64$\pm$2.68 & 4.2$\pm$0.3 \\ \hline
        9090-7 & 138.00100 & -7.84645 & 2021-01-27T04:30:36 & 56 & 0.81 & 16.8 & 43.59 & -43.17$\pm$0.69 & 12.8$\pm$0.4 \\ \hline
        F64 & 156.58800 & -0.12736 & 2021-01-27T05:18:02 & 62 & 0.73 & 16.9 & 52.22 & 29.59$\pm$0.59 & 17.1$\pm$0.4 \\ \hline
        F23 & 173.41300 & 13.44331 & 2021-01-27T06:12:20 & 79 & 0.77 & 17.0 & 60.71 & 18.92$\pm$0.47 & 24.0$\pm$0.4 \\ \hline
        CT2 & 194.38792 & -2.38783 & 2021-01-27T07:12:03 & 95 & 1.93 & 17.1 & 85.80 & 42.14$\pm$0.47 & 24.7$\pm$0.4 \\ \hline
        F74 & 235.73100 & -34.11100 & 2021-01-27T08:03:26 & 70 & 0.64 & 17.2 & 132.05 & 1.49$\pm$0.78 & 15.1$\pm$0.3 \\ \hline
        F65 & 166.39100 & -77.77269 & 2021-01-27T08:41:25 & 35 & 0.68 & 17.3 & 107.76 & 34.34$\pm$0.40 & 26.7$\pm$0.4 \\ \hline
        1101-264 & 166.20418 & -27.00247 & 2021-01-27T09:19:23 & 40 & 0.83 & 17.3 & 74.21 & -43.57$\pm$0.48 & 19.8$\pm$0.4 \\ \hline
        CT4 & 151.11239 & -2.31674 & 2017-03-12T02:26:17 & 135 & 0.94 & 18.9 & 18.81 & -39.56$\pm$3.65 & 2.3$\pm$0.4 \\ \hline
    \end{tabular}\label{Iobs-table}
\end{table*}

\end{document}